\documentclass[unsortedaddress, reprint, amsmath,amssymb,amsfonts,prb,aps, showpacs]{revtex4-1}
\usepackage{mathtools}
\usepackage{tikz}
\usepackage{tikz-qtree}
\usepackage{subfig}
\usepackage{graphicx}
\usepackage{bm}
\usepackage{amsmath}
\usepackage{multirow}
\usepackage{amssymb}

\begin{document}
	\title{Antiferromagnetic order without recourse to staggered fields}
	\author{H. C. Donker}
	\email{h.donker@science.ru.nl}
	\affiliation{Radboud University, Institute for Molecules and Materials, Heyendaalseweg 135, NL-6525AJ Nijmegen, The Netherlands}
	\author{H. De Raedt}
	\affiliation{Department of Applied Physics, Zernike Institute for Advanced  Materials, University of Groningen, Nijenborgh 4, NL-9747AG Groningen, The Netherlands}
	\author{M. I. Katsnelson}
	\affiliation{Radboud University, Institute for Molecules and Materials, Heyendaalseweg 135, NL-6525AJ Nijmegen, The Netherlands}
	
	\begin{abstract}
		In the theory of antiferromagnetism, the staggered field---an external magnetic field that alternates in sign on atomic length scales---is used to select the classical N\'eel state from a quantum magnet, but justification is missing. 
	This work examines, within the decoherence framework, whether repeated \emph{local} measurement can replace a staggered field. 
	Accordingly, the conditions under which local decoherence can be considered a continuous measurement are studied.
	The dynamics of a small magnetic system is analysed to illustrate that local decoherence can lead to (symmetry-broken) order similar to order resulting from a staggered field.
	\end{abstract}

\maketitle

\section{Introduction}
The decoherence program intends to clarify the emergence of classical physics from within quantum mechanics. 
The standard works~\cite{JOOS03, ZURE03, SCHL07} follow essentially a bottom-up approach in which the decoherence of a single particle (e.g. a quantum Brownian particle or a single spin) is discussed. Although this clarifies the fragility of quantum states of different macroscopic configurations, it does not explain how a genuine many-body system, whereby the local particles are inextricable quantum correlated (i.e., entangled), can turn into a collection of classical particles. 
Motivated by the prospects of quantum information technology, developments in the understanding of the decoherence of bi- and multipartite entangled systems are only now beginning to unfold~\cite{DUR04,CARV04,YU09,BORR09, AOLI15}. However, subtleties relating to classicality are, to the best of our knowledge, still largely unexplored.
Nevertheless, in the context of magnetism several works attempted to tackle this problem by truncating a many-body magnet to a two-level system~\cite{PROK96, PROK00,GAUY15,DELG15,DELG17}. In particular, Prokof'ev and Stamp reviewed in detail the effects of a spin environment on a two-level system~\cite{PROK96, PROK00}, for which a mapping to the spin-Boson model~\cite{LEGG87} can in general not be made. (See also a complementary review~\cite{ZHAN07} that analyses a two-level system in a spin environment from a numerical perspective.) 

On the other hand, large systems lead to emergent collective properties that are difficult to understand from a simplified sum-of-its-parts view, as discussed in several popular scientific accounts~\cite{ANDE72, LAUG00,LAUG06}. Indeed, the concept of spontaneous symmetry breaking, in which solutions are singled out in the thermodynamic limit by infinitesimal fields, depend on the (collective) low-energy behaviour of a macroscopic system~\cite{ANDE64,KOMA94}. Most systems break a(n almost) continuous symmetry---like the direction or location in space---and therefore require a host of states which conspire to form localised structures; a two-level description is then, by its very nature, inadequate.

What is more, a symmetry breaking analysis indicates \emph{if} such solutions can be singled out, but does not address the \emph{how}, i.e., the physical mechanism that is responsible. 

In magnetism, for example, this difficulty comprises the (in)consistency of the (classical) N\'eel state $|\psi_N\rangle$ (the state in which neighbouring spins align anti-parallel ${\mid\uparrow \downarrow \uparrow \dots \rangle}$ or vice versa) with the antiferromagnetic Heisenberg Hamiltonian (HH)~\cite{VONS69,VONS74}, its experimental evidence~\cite{IRKH86}, and the possible physical realisation of the staggered field (explained below). The problem is as follows: 
The exact ground state (GS) $|\psi_0\rangle$ of the antiferromagnetic HH
	\begin{equation}\label{eq:HH}
		H = J \sum_{\langle i,j\rangle} \bm{S}_i \cdot \bm{S}_j \, ,
	\end{equation}
is known to be a total spin $S_{\mathrm{tot}}=0$ singlet~\cite{LIEB62}, whereby the local magnetisation of each spin $S_i$ vanishes  $\langle \psi_0| \bm{S}_i |\psi_0\rangle = 0$ according to group theory (see, e.g., the Wigner-Eckart theorem~\cite{JONE98}). In this equation, the exchange constant $J$ is positive and the sum extends over nearest neighbours. 
Contrary to the GS, the N\'eel state $|\psi_N\rangle$ is not an eigenstate of the HH and is thus a very specific linear combination of a (possibly extensive~\cite{BERN92} number of) spin $S_{\mathrm{tot}}$ states. Moreover, the sublattice magnetisation is not a constant of motion. Therefore, the sublattice magnetisation of an arbitrary state, including the N\'eel state, will decay. 
However, since the energy levels close to the GS are very nearly degenerate, collapsing onto the GS as $1/N$ with $N$ the number of particles~\cite{ANDE52}, the decay rate can be rather long. Anderson estimated the time for the N\'eel state to rotate to an orthogonal direction to be roughly three years~\cite{ANDE52}. Furthermore, the actual GS energy $E_0$ is bounded quite strongly~\cite{ANDE51, VONS74}
	\begin{equation}\label{eq:gs_bound}
		-\frac{1}{2} NJZS_i^2 > E_0 > -\frac{1}{2} NJZS_i^2 \left[1+1/(ZS_i)\right] \, ,
	\end{equation}
with $S_i$ the single-site spin (which is taken to be the same for all $i$), and $Z$ the coordinate number of the lattice. In the limit $1/(ZS_i)\rightarrow 0$ the GS energy $E_0$ precisely coincides with the energy of the N\'eel state [left hand side Eq.~(\ref{eq:gs_bound})].

A mathematical trick to overcome the inconsistency between the N\'eel state $|\psi_N\rangle$ and the non-degenerate singlet $|\psi_0\rangle$ introduces a staggered field, $\bm{M} = \bm{S}_A - \bm{S}_B$, (the order parameter) with opposite signs on sublattice $A$ and $B$, that couples to a conjugate field, $\bm{h}_{\mathrm{st}}$.
Time-reversal invariance is thus explicitly broken by adding the term $H_{\mathrm{st}} = \bm{M}\cdot \bm{h}_{\mathrm{st}} $ to the Hamiltonian. 
The sublattice magnetisation then arises as a quasi-average~\cite{BOGO60}, whereby the conjugate field $\bm{h}_{\mathrm{st}}$ tends to zero after taking the thermodynamic limit. 
Although it is known that an effective staggered field can be generated in very specific crystal structures~\cite{OSHI97}, it is usually regarded as unphysical~\cite{ZIMA52, IRKH86, KOMA94, KUZE10}. More generally, the proper choice of the order parameter can not always be decided \emph{a priori}, but is dictated by phenomenology~\cite{HUAN00}.

Assume now, for the sake of argument, that the staggered field does have a physical origin.
For natural occurring magnetic fields of arbitrary shape that are infinitely differentiable, the Fourier component corresponding to the staggered field is suppressed faster than any power of the spectral scale (which is typically much larger than the lattice spacing) and hence decreases super exponentially below this scale. Therefore, arbitrary stray fields emanating from outside into the sample are an unlikely source of the staggered field.

In fact, for the interpretation of neutron diffraction experiments no such field is required~\cite{PRAT61, IRKH86, LHUI05}. 
Elastic neutron diffraction experiments probe the time-reversal invariant static structure factor [Eq.~(\ref{eq:Kaa})], and therefore do not require symmetry-broken states~\cite{PRAT61, IRKH86, LHUI05}. 
An explicit demonstration was given by Irkhin and Katsnelson~\cite{IRKH86}, who proposed a trial wave function \emph{without} broken symmetry for the Heisenberg antiferromagnetic model in the semi-classical $1/(ZS_i) \rightarrow 0$ limit. Not only were they able to reproduce the peaks at the antiferromagnetic reciprocal lattice vector in the static structure factor. Also the nuclear magnetic resonance line form could be accounted for without resorting to broken symmetry.

Besides the existence and necessity of the  staggered field, it is \emph{a priori} unclear whether this mathematical trick indeed leads to non-vanishing anomalous averages $\langle \bm{M}\rangle \neq 0$ upon sending $\bm{h}_{\mathrm{st}} \rightarrow 0$ after taking the thermodynamic limit. The Lieb-Mattis model~\cite{LIEB62} is one of the few non-trivial systems where this can be worked out in detail~\cite{KAIS89,KAPL90}.
More generally, the ability to develop spontaneous staggered magnetisation hinges on the presence of long-range order.
In the antiferromagnetic Heisenberg model on the cubic lattice, long-range order was first established by Dyson, Lieb, and Simon~\cite{DYSO78} for $d \geq 3$ spatial dimensions and spin $S_i \geq 1$. This proof was later strengthened by others for $d=3$ to include the case $S_i=1/2$~\cite{KENN88}. 
In two spatial dimensions, numerical evidence suggest long-range order in the GS for both the bipartite~\cite{LHUI05,LOW07,WHIT07} and the triangular lattice~\cite{JOLI90,BERN94,CAPR99,WHIT07} when $S_i=1/2$ (for other lattices see also~\cite{LHUI05} and references therein).
But only fairly recently it was demonstrated that taking an anomalous average does indeed lead to broken symmetry for the HH on a bipartite lattice, provided that long-range order exists~\cite{KAPL89,KOMA93}.

Later refinements of several experimental techniques called for a revaluation of the magnetic ordering in antiferromagnets. Measurements of the quadrupole magnetic moment of the antiferromagnetic compound Cr$_2$O$_3$ indicated broken symmetry in the magnetic structure~\cite{ASTR96}. Furthermore, state-of-the-art spin-polarised scanning tunnelling microscopy experiments can probe the magnetisation of individual atoms and have revealed antiferromagnetic structures whereby time-reversal symmetry is manifestly broken. Moreover, these structures have been seen to telegraph between the two alternating N\'eel configurations~\cite{LOTH12,YAN17}, reminiscent of Bohr's quantum jumps in atoms~\cite{JOOS03}. (In these works~\cite{LOTH12,YAN17}, they considered $S=2$ Fe atoms on Cu$_2$N that have a large magnetic easy-axis anisotropy~\cite{HIRJ07} which consequently suppresses quantum fluctuations.) 
This has sparked renewed interest how such classical magnetic order can come about from quantum systems~\cite{GAUY15,DELG15,DELG17}. 
This work contributes to the discussion by analysing the necessity of the staggered field, going beyond a two-level simplification.

Historically, the decoherence program focused primarily on single-particle subsystems but, as it turned out, the decoherence of local particles in a many-body subsystem can lead to many surprising consequences. 
Examples are the decoherence wave~\cite{KATS00,HAMI05,DONK16}, suppression of the Kondo effect~\cite{KATS03,KATS05}, and the creation of sublattices in antiferromagnets~\cite{KATS01}, all of which considered idealised local measurement (i.e., a wave function collapse). 
This work follows up on a suggestion in Ref.~\cite{KATS01} that repeated local measurements on antiferromagnets can replace a staggered field.  
A decoherence approach is followed, in which the entanglement between a local spin of an antiferromagnet and an environment is studied. Since the decoherence process is technically different from wave function collapse, first a detailed comparison is carried out.
Next, a critical assessment is made to clarify under which conditions, and to what extent, the repeated measurement hypothesis is correct with the help of numerical calculations. 

For clarity, the analysis is broken up in two parts.
In model $A$, the decoherence of a local particle is compared to idealised measurement (Sec.~\ref{sec:local_DW}). In model $B$, local decoherence is applied  to a low-energy description of antiferromagnets (Sec.~\ref{sec:staggered_field}). 
Both models are outlined in Sec.~\ref{sec:model} and  implications are discussed in Sec.\ref{sec:discussion}.

\section{Outline models}\label{sec:model}
The two spin models that are the centre piece of this work shall now be outlined. Units in which the reduced Planck constant $\hbar$ and the Boltzmann constant $k_B$ equal unity are used throughout this paper.
In both models, the entire system consists of a collection of spin-1/2 particles. A single (local) particle is strongly coupled to an environment along the $z$-direction, so as to resemble a measurement-like interaction. 
While model $A$ mainly serves to compare ideal measurement with local decoherence, $B$ extents the environment of $A$ by introducing a thermal reservoir. An important difference between $A$ and $B$ is that in the latter an effective description of antiferromagnets is used (for a detailed discussion see Sec.~\ref{sec:staggered_field}) for the system of interest [henceforth central system (CS)].

\subsection{Model $A$: Decoherence of a local spin}\label{sec:model_A}
\begin{table}
\centering
	\begin{tabular}{l | l}
	Model $A$ & Model $B$\\
	\hline
	$N_S=6$ & $N_S=4$\\
	\multirow{2}{*}{$N_{\cal E}=8$} & $N_{{\cal E}_1}=6$ \\
	 & $N_{{\cal E}_2}=12$\\
	 $I = 20 J_S$ & 	 $I = 20 J_S$ \\
	 & $I^\prime = 0.1 J_S$\\
	 & $K=0.1J_S$\\
	 & $\beta = 50/J_S$
	\end{tabular}
	\caption{Model parameters; interaction strengths are expressed in units relative to the exchange constant, $J_S$, of the central system.}
	\label{tab:param}
\end{table}
The entire system is partitioned in two parts, the CS (denoted by $S$) and its complement, the environment (indicated by ${\cal E}$). The CS (environment) is composed of $N_S=6$ ($N_{\cal E}=8$) spin-1/2 particles. 
To avoid unnecessarily complicating the analysis, the CS with Hamiltonian $H_S^{\text{---}}$ interacts locally with environment ${\cal E}$ via $H_I$, 
while the spins in ${\cal E}$ have no Hamiltonian of their own. That is, the intra-environment Hamiltonian  (the self-Hamiltonian of ${\cal E}$) is neglected.
The Hamiltonian of model $A$ can then be written as
\begin{equation}
H_A = H_S^{\text{---}} + H_I \, .
\end{equation}
The CS is taken to be an open chain of $N_S$ spins (i.e., $S_i=1/2$ for all $i$), coupled via Heisenberg exchange~\cite{VONS74}
\begin{equation}\label{eq:HH_obc}
H_S^{\text{---}} = J_S \sum_{i=1}^{N_S-1} \bm{S}_i \cdot \bm{S}_{i+1} \, ,
\end{equation}
with exchange constant $J_S$, and the "---" superscript indicating the lattice geometry (in this case an open chain). To emulate decoherence as a \emph{measuring} effect on a local spin, which is chosen to be spin $S_1$, requires a well defined measurement direction. The $z$-axis is selected, such that the system-environment interaction takes the following form
\begin{equation}\label{eq:HI}
H_I = I \sum_{i \in {\cal E}} r_i S_1^z S_i^z \equiv  I S_1^z \tilde{S}_{\cal E}^z \, ,
\end{equation}
where the sum is over all $N_{\cal E}$ spins inside the environment ${\cal E}$, $I$ is the interaction strength, $\{r_i\}$ are a set of random numbers $r_i \in [0,1]$, and $\tilde{S}_{\cal E}^z = \sum_{i \in {\cal E}} r_i S_i^z$. The use of random numbers are to suppress recurrences of phase coherence after the initial Gaussian decay. In order for ${\cal E}$ to couple strongly to the CS, $I$ is set to $I=20J_S$. 

\subsection{Model $B$: Addition of a thermal reservoir}\label{sec:model_B}
\begin{figure}
	\centering
	{\includegraphics[scale=0.15]{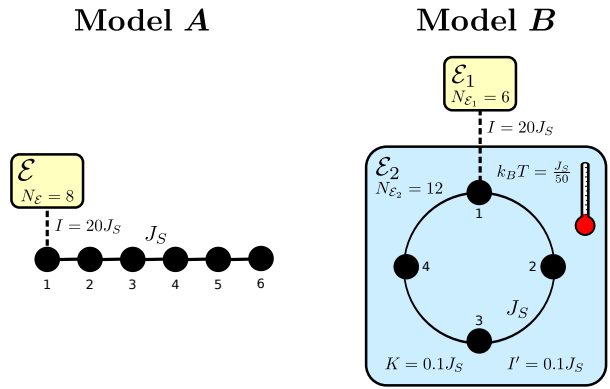}}
	\caption{Schematic of models $A$ and $B$. In model $A$ ($B$) the central system consists of an open (periodic) chain of $N_S=6$ ($N_S=4$) spin-1/2 particles and spin $S_1$ is strongly coupled to an environment ${\cal E}$ (environment fragment ${\cal E}_1$) of spins via Ising coupling.	
In model $B$ the spin chain is immersed in fragment ${\cal E}_2$ that resembles a thermal reservoir (inverse temperature $\beta=50/J_S$). The number of spins in each environment (fragment) and the respective coupling strengths are indicated in the figure.}
	\label{fig:model_B}
\end{figure}
In model $B$, the CS interacts with an environment $\cal E$, but in this case ${\cal E}$ is consists of two fragments ${\cal E}_1$ and ${\cal E}_2$. Fragment ${\cal E}_1$ describes the strong decoherence with the local spin, while ${\cal E}_2$ couples weakly to the entire CS to mimic contact with a thermal reservoir.
The Hamiltonian of model $B$ is split into four: 
\begin{equation}
H_B = H_S^{\bigcirc} + H_{I_1} + H_{I_2} + H_{{\cal E}_2} \, ,
\end{equation}
with $H_S^{\bigcirc}$ the Hamiltonian of the CS (the $\bigcirc$ indicates ring geometry), $H_{I_1}$ ($H_{I_2}$) the coupling between the CS and ${\cal E}_1$ (${\cal E}_2$), and $H_{{\cal E}_2}$ the self-Hamiltonian (i.e., the intra-environment Hamiltonian) of ${\cal E}_2$. A schematic of the set-up is shown in Fig.~\ref{fig:model_B}. The Hamiltonian, $H_S^{\bigcirc}$, is given by Eq.~(\ref{eq:HH_obc}) with the addition of the boundary term $H_S^{\bigcirc} = H_S^{\text{---}} + J_S \bm{S}_1 \cdot \bm{S}_{N_S}$, thereby giving rise to a ring geometry.

Similar to model $A$, spin $S_1$ of the CS is strongly coupled to spins in ${\cal E}_1$ via Ising coupling. $H_{I_1}$ is identical to Eq.~(\ref{eq:HI}), apart from the sum that is now restricted to the $N_{{\cal E}_1}$ sites pertaining to fragment ${\cal E}_1$. 

For ${\cal E}_2$, slightly different system-environment couplings are used to facilitate decoherence of the energy states. In Ref.~\cite{YUAN07} it was found that binary coupling strengths and the presence of a conserved quantity are particularly efficient. 
Therefore, the following coupling to ${\cal E}_2$ is chosen
\begin{equation}
H_{I_2} = \sum_{i \in S, k \in {\cal E}_2 } I_{ik}^{\prime} S^z_i S^z_k \, ,
\end{equation}
where $i$ runs over spin indices in the CS (indicated by $S$) and $I_{ik}^{\prime}$ are  binary values $\pm I^\prime$ picked at random.
As for the self-Hamiltonian of ${\cal E}_2$, spin-glass like couplings are used to maximise decoherence and relaxation~\cite{YUAN07,YUAN06}
\begin{equation}
H_{{\cal E}_2} = \sum_{\alpha \in \{x,y,z\}} \sum_{k,l \in {\cal E}_2} K_{kl}^\alpha S_k^\alpha S_l^\alpha \, ,
\end{equation}
with $K_{lm}^\alpha$ uniform random numbers in the range $[-K,K]$. 
The philosophy behind the specific form $H_{{\cal E}_2}$ is that it is not necessary to have a very large environment in order to have efficient decoherence and relaxation~\cite{YUAN07,YUAN06}, thereby keeping the problem computationally tractable. But this comes at the expense of having to choose specific---namely, spin-glass---types of couplings for the bath.

In order to prevent energy flow from ${\cal E}_2$ into the CS, the environment is prepared in a configuration that resembles a thermal state with inverse temperature $\beta = 50/J_S$, i.e., very close to the GS. Both decoherence and relaxation are sensitive to the precise numerical values of the interaction strengths. The values that are picked lead to efficient decoherence and relaxation for the given size of the environment. The interaction strengths of both models (as well as other parameters) are summarised in Table~\ref{tab:param}.

\subsection{State preparation and simulation procedure}
In order to study the decoherence process, it is most instructive to examine a state that is initially unentangled, i.e., a product state. In particular, this work shall encompass the decoherence of a CS that is prepared in the GS, $|\psi_0\rangle$, (respective to the model) at time $t=0$. The global wave function, $|\Psi(t)\rangle$, is then expressed as
\begin{equation}
|\Psi(0)\rangle = |\psi_0\rangle \otimes |{\cal E}_0\rangle \, ,
\end{equation}
with $|{\cal E}_0\rangle$ the initial state of ${\cal E}$. The state $|{\cal E}_0\rangle$ is constructed using the Box-Muller method~\cite{BOX58} to generate a random state. In the case of model $B$, an additional step is required to turn fragment ${\cal E}_2$ into a thermal-like state. This is done by performing imaginary time evolution $\exp[-\beta H_{{\cal E}_2}/2]$ on the random state of ${\cal E}_2$~\cite{HAMS00} and subsequent normalisation of the resulting wave function. 

Time evolution of the global state $|\Psi(t)\rangle$ in model $A$ ($B$) is governed by the unitary operator $\exp[-itH_{A}]$ ($\exp[-itH_{B}]$). Time $t$ shall consistently be expressed in dimensionless form $t=t^\prime J_S/\hbar$, in which $t^\prime$ is dimensionful time and $\hbar$ was restored for clarity. Both real and imaginary time evolution are numerically calculated by expanding the exponential in Chebyshev polynomials~\cite{DOBR03,RAED06}. The expansion allows for calculation of the wave function with an accuracy up to machine precision~\cite{DOBR03,RAED06}. This accuracy is important to unambiguously assign loss of phase coherence to quantum entanglement instead of the accumulation of numerical errors. 

Finally, the loss of phase coherence is studied by taking partial traces of the density matrix~\cite{FANO57} of the global system $\Pi(t) = |\Psi(t)\rangle \langle\Psi(t)|$. The resulting reduced density matrix (RDM) of the CS is defined as
\begin{equation}\label{eq:rho_red}
\rho(t) = \mathrm{Tr}_{\cal E} \Pi(t) \, ,
\end{equation}
where the trace is over all the spin states in ${\cal E}$.
For each simulation a new realisation of the environment was generated as well as a new set of random couplings.
The data shown here is representative for simulations with different random realisations. 

\section{Comparison of decoherence with the quantum Zeno effect}\label{sec:local_DW}
\begin{figure}
	\centering
	\includegraphics[scale=0.25]{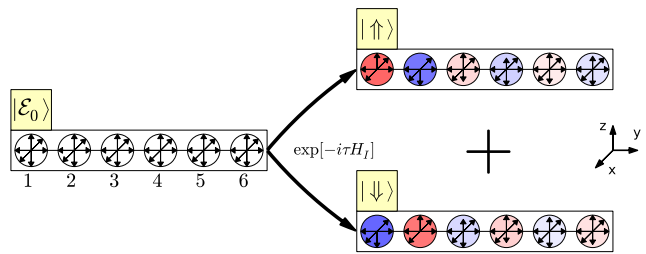}
	\caption{Pictorial representation of the initial decoherence process in model $A$. Different parts of the (initial) singlet state evolve to (approximately) orthogonal parts of the environment (denoted by $\mid \Uparrow\rangle$ and $\mid \Downarrow\rangle$) via spin $S_1$. The boxes represent a product state of the central system with the environment. 
The magnetisation of each spin is indicated by the arrows, and the red (blue) colour intensity illustrates the net magnetisation parallel (anti-parallel) to the z-axis.}
	\label{fig:singlet_branch}
\end{figure}
As stated in the Introduction, the primary goal is to understand the consequences of repeated local measurement---the local analogue of the quantum Zeno effect~\cite{MISR77, JOOS03}---on antiferromagnets. This section examines to what extent repeated local measurement can be described within the framework of decoherence.

It is intuitively clear that, from the decoherence perspective, a continuous measurement might originate from an environment interacting much longer than the typical time scale of the CS. One key requirement is that coherences can be quenched locally while maintaining global coherence in the CS. 
To make the connection between decoherence and the quantum Zeno effect explicit, let  us now turn to model A (see Sec.~\ref{sec:model_A}) and invoke the Trotter-Suzuki product formula~\cite{SUZU76}: 
\begin{equation} \label{eq:trotter}
\exp[-it H_A] \approx [ \exp(-itH_S^{\text{---}}/n) \exp(-itH_I/n)]^n \, ,
\end{equation}
in which the approximation becomes exact if $n$ tends to infinity.
Under time evolution the wave function initially branches (partially) due to $\exp[-itH_I/n]$, as different parts of the CS singlet state, $|\psi_0\rangle$, entangle to mutually (close to) orthogonal environment states (see Fig.~\ref{fig:singlet_branch}). Subsequently, each branch evolves individually for a time $t/n$ under $H_S^{\text{---}}$. In comparison, the time evolution according to the quantum Zeno effect is governed by the operator $T_n(t) = [\exp(-itH_S^{\text{---}}/n) P_1^\pm]^n$ with $P_i^\pm = [1 \pm \sigma_i^z]/2$ the spin $S_i$ projection operator. In the limit where the decoherence time scale $\tau$ (the time scale that makes the relative states of ${\cal E}$ orthogonal) tends to zero and $n$ tends to infinity, the descriptions $\exp[-it H_A]$ [see Eq.~(\ref{eq:trotter})] and $T_n(t)$ become compatible. In this limit the repeated application of this two-step process causes spin $S_1$ to be pinned, while the remaining spins in each branch are allowed to evolve freely under $H_S^{\text{---}}$.

In physically more realistic systems, these precise mathematical limits are never reached. As a result, $\tau$ stays finite but small. Therefore, if the Trotter-Suzuki product formula---which has a structure similar to the quantum Zeno time-evolution operator $T_n(t)$---is used to approximate the evolution operator $\exp[-itH_A]$ with finite $\tau = t/n$, then the contributions coming from the commutator $[H_S^{\text{---}}, H_I]$ as well as higher order commutators are neglected, as can be made explicit using the Campbell-Baker-Hausdorff formula~\cite{STERN04}.

\subsection{Results model $A$}
\begin{figure*}
	\centering
	\subfloat[Coherence local to spin $S_1$ {[dashed line, Eq.~(\ref{eq:coh_loc})]} and the remainder of the subsystem {[solid line, Eq.~(\ref{eq:coh_glb})]}, as a function of time. Coherence is with respect to the computational basis.]{\label{fig:local_coh}
	\includegraphics[scale=0.27]{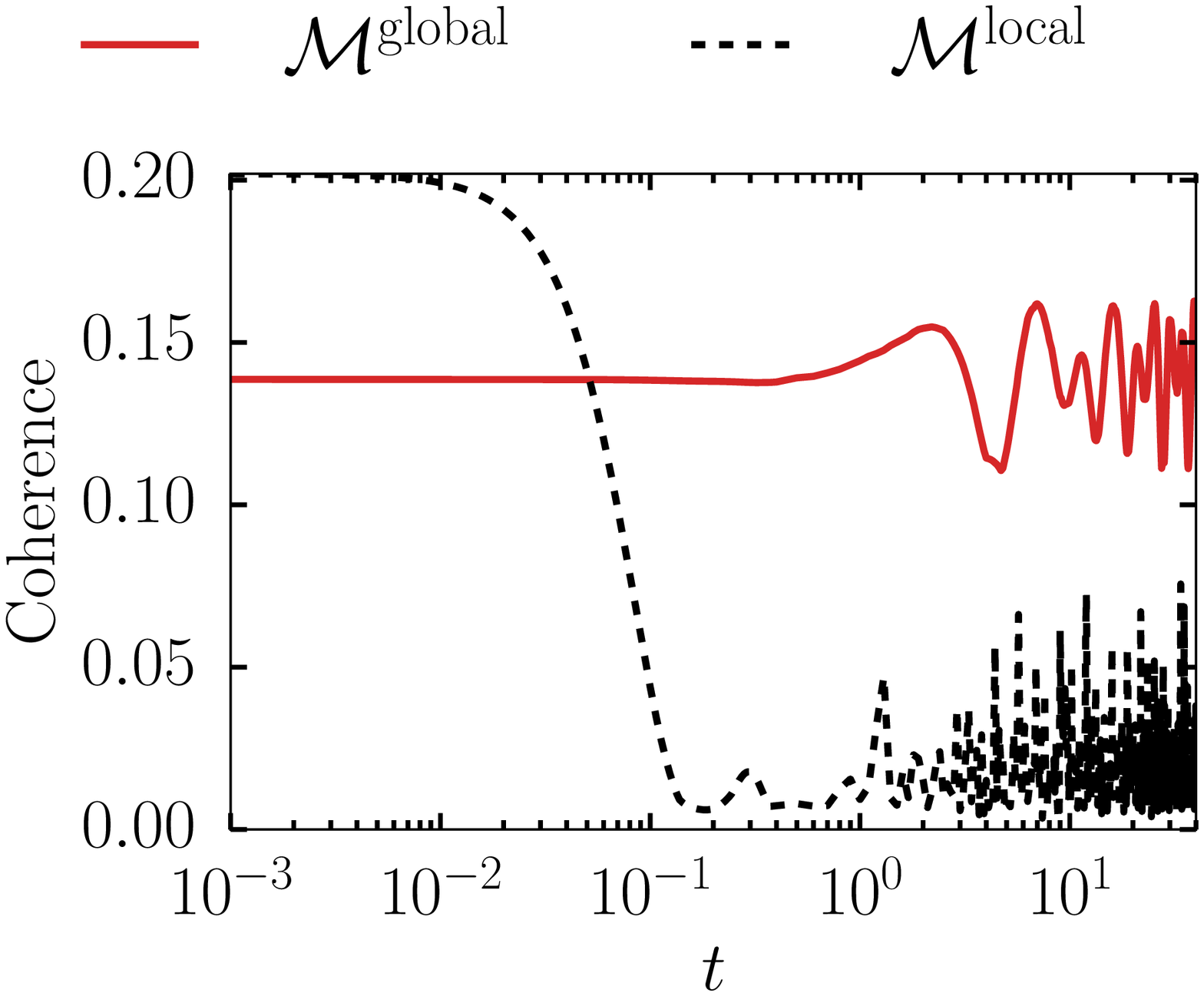}}
	\qquad
	\subfloat[Entropy of the entire subsystem {$S(t) = S[\rho(t)]$} (solid line), and that complementary to spin $S_1$, {$S_{\uparrow\uparrow}(t) = S[\tilde{\rho}_{\uparrow\uparrow}(t)]$} (dashed line) and {$S_{\downarrow\downarrow}(t) = S[\tilde{\rho}_{\downarrow\downarrow}(t)]$} (dotted line). 
	]{\label{fig:local_entr}
	\hfill
	\includegraphics[scale=0.27]{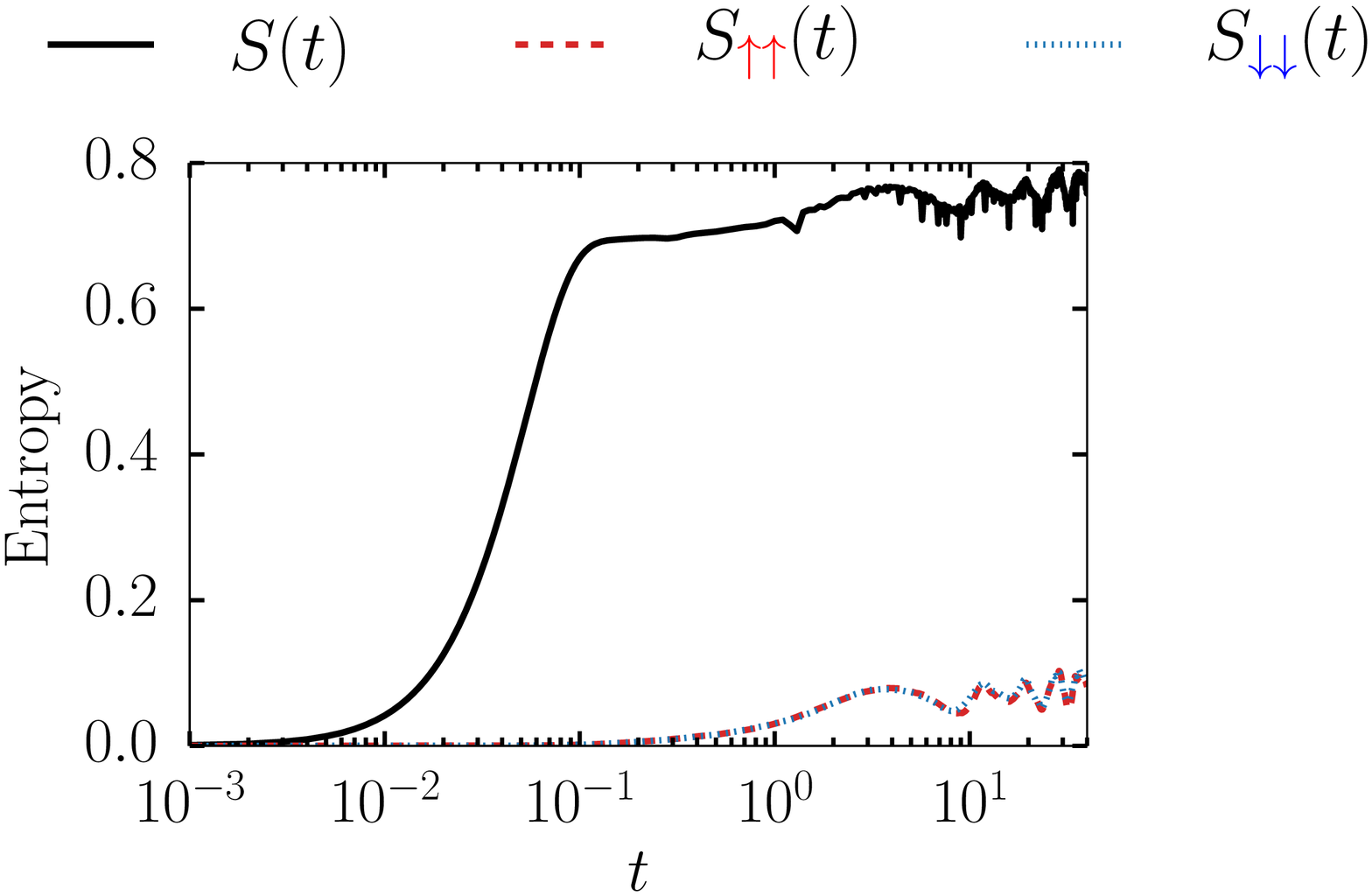}}
	\caption{Simulation results of model $A$, 
	where spin $S_1$ of a $N_S=6$ Heisenberg (open) spin chain is coupled to $N_{\cal E}=8$ environment spins via Ising coupling with random uniform interaction strength $I=20J_S$. 
	The central system is prepared in the (singlet) ground state at $t=0$.
	}
	\label{fig:HH_simple_decoh}
\end{figure*}

The simulation results of model $A$ will now be compared to the quantum Zeno picture, and it will be verified that local decoherence leads to a decoherence wave (DW).  
The representation $|i_1 \dots i_{N_S}\rangle$ in which each $i_k$ takes the value $\uparrow$ or $\downarrow$ shall henceforth be referred to as the computational basis.

The preceding discussion (and as illustrated in Fig.~\ref{fig:singlet_branch}) suggests that coherence in the computational basis between \makebox{$\mid \uparrow, i_2,\dots,i_{N_S}\rangle$} and \makebox{$\mid \downarrow, j_2,\dots,j_{N_S}\rangle$} diminishes for all realisations of the undetermined indices. 
At the same time, states with identical $S_1^z$ eigenvalues are expected to be unaffected in the quantum Zeno limit. 
Therefore, to quantify the local coherence of the RDM
	\begin{equation}
		\rho_{\{ i_1 i_2 i_3 \dots| j_1 j_2 j_3 \dots \}} \equiv \langle i_1, i_2, i_3\dots\mid \rho\mid j_1, j_2, j_3 \dots \rangle 
	\end{equation}
it is beneficial to focus on the $i_1$ and $j_1$ components.
To this end, consider the RDM conditioned on the spin $S_1$ components (as indicated by the bra/ket subscript)
	\begin{equation}
		\rho_{ab} = {}_1\langle a | \rho | b \rangle_1 \, ; \quad \tilde{\rho}_{ab} = \rho_{ab} / \mathrm{Tr}[\rho_{ab}] \, .
	\end{equation}
Normalisation of the density operator $\tilde{\rho}_{ab}$ is primarily to compare entropy, as discussed below.
The matrix elements of $\rho_{\uparrow \downarrow}$ (and $\rho_{\downarrow \uparrow}$) determine the degree of (local) $S_1$ coherence.
To measure the loss of local coherence, for each time step $t$, the maximal magnitude (absolute value) of the $|\rho_{\uparrow \downarrow}|$ components
	\begin{equation}\label{eq:coh_loc}
		{\cal M}^{\mathrm{local}}(t) = \mathrm{max}_{i,j}\left[|\langle i|\rho_{\uparrow \downarrow}(t)|j\rangle|\right] \, ,
	\end{equation}
is calculated, with $|i\rangle = |i_2, i_3,\dots,i_{N_S}\rangle$ the remaining spins evaluated in the computational basis, and likewise for $|j\rangle$. The coherence of the rest of the CS is determined by the off-diagonal components of $|\rho_{\uparrow \uparrow}|$ and $|\rho_{\downarrow \downarrow}|$, which can similarly be quantified as
	\begin{equation}\label{eq:coh_glb}
	{\cal M}^{\mathrm{global}}(t) = \mathrm{max}_{i \neq j}\left[|\langle i|\rho_{\uparrow \uparrow}(t)|j\rangle|\right] \, ,
	\end{equation}
where $i$ and $j$ like above, which do not coincide $i\neq j$ (it must be an off-diagonal component of $\rho$).
The time evolution of the components are shown in Fig.~\ref{fig:local_coh}.
 
Two regions can be identified: 1) the dephasing regime with $t < 1$ and; 2) the dynamic regime $t \sim 1$. Fig.~\ref{fig:local_coh} illustrates that in region 1) the global coherence in the CS is essentially unperturbed whilst the local coherence associated with spin $S_1$ is suppressed.
Region 2) is determined by the Hamiltonian $H_S^{\text{---}}$, and it is on this time scale that the DW manifests itself.
The oscillatory behaviour of ${\cal M}^{\mathrm{local}}(t)$ starting $t \sim 10^0$ are recurrences that originate from the finite size of $N_{\cal E}$; additional suppression can be achieved by increasing $N_{\cal E}$.

To further quantify the system's coherence, it is helpful to introduce the von Neumann entropy~\cite{NEUM55}
\begin{equation}
S[\rho(t)] = -\mathrm{Tr}[\rho (t) \ln \rho(t)] \, ,
\end{equation}
which measures the purity of the density matrix $\rho(t)$ (it vanishes for pure states).
Analogously, the entropy of the (normalised) spin $S_1$ diagonal components of the RDM are defined as $S_{nn}(t) = S[\tilde{\rho}_{nn}(t)]$. 
In the quantum Zeno description, the diagonal components $\tilde{\rho}_{nn}$ are by definition pure (since it describes wave function collapse, leading to a new pure state). By the normalisation of $\tilde{\rho}_{nn}$, this would imply $S_{nn}(t)=0$.
The entropy of the RDM $S(t) \equiv S[\rho(t)]$ as well as $S_{\uparrow \uparrow}(t)$ and $S_{\downarrow \downarrow}(t)$, are shown in Fig.~\ref{fig:local_entr}.
 What can be seen is that, up to $t \approx 1$, the increase in entropy of $\rho(t)$ (tending towards $S = \ln 2 \approx 0.69$) can be primarily attributed to the decoherence of spin $S_1$ (see also Fig.~\ref{fig:local_coh}). For larger times, $t$, the coherence of the CS is somewhat diminished on a more global scale as the entropy of the spin $S_1$ diagonal components increase. In addition, the entropy of $\tilde{\rho}_{\uparrow\uparrow}$ is almost the same as that of $\tilde{\rho}_{\downarrow\downarrow}$. This can be understood by noting that, from the local spin perspective, the random environment state looks similar when all the spins are reversed. Simulations, not shown here, indeed indicate that the entropy difference between the two diagonal components varies for each random realisation of the environment.

\begin{figure}
\centering
\includegraphics[scale=0.33]{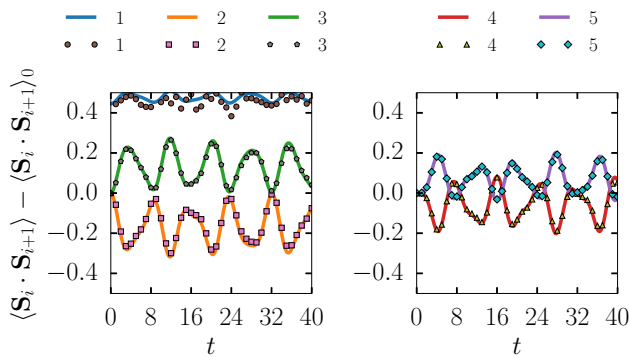}
\caption{ 
Nearest-neighbour correlations $\langle \bm{S}_i(t) \cdot \bm{S}_{i+1}(t)\rangle$ of a $N_S=6$ antiferromagnetic open Heisenberg chain for different sites $i$.
Repeated collapse (solid lines) along the $z$-direction of spin $S_1$ (performed every $\Delta t = 10^{-1}$ units of time) is compared with decoherence model $A$ (markers).
For clarity, the correlations $\langle \bm{S}_i(t) \cdot \bm{S}_{i+1}(t) \rangle$ with $i=1,2,3$ ($i=4,5$) are shown in the left (right) panel and the $t=0$ ground state values, $\langle \bm{S}_i(0) \cdot \bm{S}_{i+1}(0)\rangle \equiv \langle \bm{S}_i \cdot \bm{S}_{i+1} \rangle_0$, have been subtracted for each $i$.
}
\label{fig:ei}
\end{figure}

To see how the non-ideal aspect, whereby decoherence not only affects spin $S_1$ but also the remainder of the system, modifies the DW, consider now a time-reversal invariant observable, such as the local energy of the Heisenberg spin chain $\langle \bm{S}_i \cdot \bm{S}_{i+1} \rangle$. 

In Fig.~\ref{fig:ei} the nearest-neighbour correlations from the decoherence process are compared to repeated (every $\Delta t = 10^{-1}$) collapse of spin $S_1$ along the $z$-direction. The latter is achieved by applying the site $i$ projection operator $P_i^{\pm} = [1 \pm \sigma_i^z]/2$ to the wave function. Although $\langle \bm{S}_1 \cdot \bm{S}_{2} \rangle$ (left panel Fig.~\ref{fig:ei}) shows some deviation between decoherence (circular markers) and collapse (thick blue line), the other sites show very good quantitative agreement. The apparent scattering of the $\langle \bm{S}_1 \cdot \bm{S}_{2} \rangle$ markers actually originates from fast oscillatory behaviour. To stress, the data points of both the markers and the solid lines are solutions to the Schr\"odinger equation that are numerically accurate up to machine precision. The deviation between the solutions can therefore be solely attributed to the degree to which the two descriptions are compatible. Further agreement can be achieved by increasing the interaction strength, thereby decreasing the relative importance of the non-commutative contributions, and increasing $N_{\cal E}$ to negate finite-size effects. 

It can thus be concluded that, even though the description of the decoherence process in terms of the quantum Zeno effect (meaning: repeated wave function collapse) is approximate, in practice the two descriptions show a fair degree of compatibility. 

\section{Staggered field from the quantum Zeno effect}\label{sec:staggered_field}
Let us start by discussing the hierarchy of the low-lying excitations in magnetic systems~\cite{BERN92, BERN94, KOMA94, CAPR01, LHUI05, WEZE06}. The basic tenet is that, in the thermodynamic limit, the collective dynamics (of the antiferromagnet as a whole) are slow compared to the time scale pertaining to the internal excitations that describe local modulations of magnetic order~\cite{ANDE64}. Hence, the collective configuration---such as absolute position of a crystal or sublattice magnetisation direction in an antiferromagnet---can be presumed fixed in comparison to the time interval wherein internal dynamics are relevant~\cite{ANDE64}. This, of course, still requires that the initial state of the system has a well defined collective configuration to begin with. 

To further discuss the ordering of energy levels, consider the HH [Eq.~(\ref{eq:HH})] in Fourier space (Latin and Greek indices refer to real and Fourier space, respectively) in $d$ spatial dimensions
\begin{equation}\label{eq:HH_fourier}
H = J\sum_{\bm{\kappa}} \gamma_{\bm{\kappa}} \bm{S}_{\bm{\kappa}} \cdot \bm{S}_{-\bm{\kappa}} = H_{\mathrm{LM}} + J\sum_{\bm{\kappa} \neq 0, \pi} \gamma_{\bm{\kappa}} \bm{S}_{\bm{\kappa}} \cdot \bm{S}_{-\bm{\kappa}} \, ,
\end{equation}
with $\gamma_{\bm{\kappa}} =  \sum_i \cos(\bm{\kappa} \cdot \bm{u}_i) $ a sum over primitive vectors $\bm{u}_i$, and where
\begin{equation}
\bm{S}_{\bm{\kappa}} = \frac{1}{\sqrt{N}} \sum_l e^{i\bm{\kappa}\cdot \bm{R}_l} \bm{S}_l \, ,
\end{equation}
defines the Fourier transform of spin operators $\bm{S}_l$, and $\bm{R}_l$ the respective lattice positions (in units of lattice spacing) with periodic boundary conditions. On the right hand side of Eq.~(\ref{eq:HH_fourier}) the $\bm{\kappa}=0$ and $\bm{\kappa}=\pi$ contribution are separated to form $H_{\mathrm{LM}}$. On a bipartite lattice $H_{\mathrm{LM}}$ turns out to be the Lieb-Mattis~\cite{LIEB62} Hamiltonian~\cite{BERN92, BERN94, CAPR01, WEZE06,LHUI05}
\begin{equation}\label{eq:lm}
H_{\mathrm{LM}} = \frac{4dJ}{N} \sum_{i=1, j=1}^{N/2} \bm{S}_{2i-1} \cdot \bm{S}_{2j}  = \frac{J^\prime}{N} \bm{S}_A \cdot \bm{S}_B \, ,
\end{equation}
whereby the odd (even) sites refer to sublattice $A$ ($B$) and $J^\prime \equiv 4dJ$. For this system, the lowest energy levels are total spin $S_{\mathrm{tot}}$ states with maximal $S_A$ and $S_B$ that collapse onto the GS as $ J^\prime S_{\mathrm{tot}}(S_{\mathrm{tot}}+1)/N$~\cite{KAIS89,KAPL90}. To compare this to the dynamics of $H$, the complement of $H_{\mathrm{LM}}$ can be treated in linear spin wave theory~\cite{ANDE52}. The "softest" magnon is separated from its ground state as $\propto J/N^{1/d}$, thereby justifying the hierarchy in time scales---in which the collective dynamics are slow compared to the internal magnon excitations---for $d>1$ (and not too large $S_{\mathrm{tot}}$) in the spin-wave picture~\cite{BERN92, BERN94, CAPR01, LHUI05}. 

\emph{A posteriori} analysis of the energy levels of specific systems indicate that the aforementioned dichotomy between collective and local dynamics can indeed be found in many antiferromagnetic systems for $d=2$ dimensions~\cite{BERN94, FOUE01,LHUI05}. Not only in near-neighbour antiferromagnets on various lattices~\cite{BERN92, BERN94} but also in antiferromagnetic systems with further-neighbour interactions~\cite{FOUE01,LHUI05}. This, then, is another example whereby the emergent physical state is insensitive to the precise microscopic details of the Hamiltonian, as discussed by Laughlin and Pines~\cite{LAUG00}.
Having established that the lowest-lying energy levels of the Lieb-Mattis model approximately describe the respective states of the Heisenberg Hamiltonian (for various lattices and geometries), let us proceed to discuss the consequences of local decoherence.
Assume now, for simplicity, that all individual spins (i.e., for both sublattices) are $S_i=1/2$ and consider the decoherence of spin $S_1$ positioned on sublattice $A$. 
To reiterate, the decoherence of a spin and the quantum Zeno effect are strictly speaking inequivalent. But as demonstrated in Sec.~\ref{sec:local_DW}, in practice the two descriptions are to a large degree compatible. Therefore, assume that spin $S_1$ is decohered sufficiently strong such that it is, for all practical purposes, pinned along the $z$-direction by its environment.

The branch of the wave function corresponding to spin up (spin down) can now be described by an effective Hamiltonian $H_{\mathrm{eff}}^+$ ($H_{\mathrm{eff}}^-$):
\begin{equation}
H_{\mathrm{eff}}^\pm = \frac{J^\prime}{N} \bm{S}_{A^\prime} \cdot \bm{S}_B \pm \frac{J^\prime}{2N} S_B^z \, ,
\end{equation}
whereby the spin of the reduced sublattice $\bm{S}_{A^\prime} = \sum_{i=2}^{N/2} \bm{S}_i$ was introduced. If the initial state of the CS was the GS, belonging to the $S^z_{\mathrm{tot}}=S_A^z + S_B^z=0$ subspace, the effective Hamiltonian can be cast in the following suggestive form
\begin{equation}\label{eq:HLM_eff}
{H_{\mathrm{eff}}^\prime}^\pm = \frac{J^\prime}{N} \bm{S}_{A^\prime} \cdot \bm{S}_B \mp h_{\mathrm{st}}(S_{A^\prime}^z - S_B^z) \, , 
\end{equation}
with $h_{\mathrm{st}}=J^\prime/(4N)$, whereby an additive constant due to spin $S_1$ was dropped. Note that in the decoherence framework, time-reversal symmetry is manifestly preserved; the global state of the CS plus environment describes a superposition of two reduced CS (spin $S_1$ no longer partakes in any dynamics), subject to equal but opposite staggered fields. 

\section{Demonstration emergent magnetic order by decoherence}
\begin{figure*}
	\subfloat[Simulation data of the static structure factor evaluated at the antiferromagnetic reciprocal lattice vector (i.e., $\kappa=\pi$) as a function of time.]{\label{fig:Kt}\includegraphics[scale=0.3]{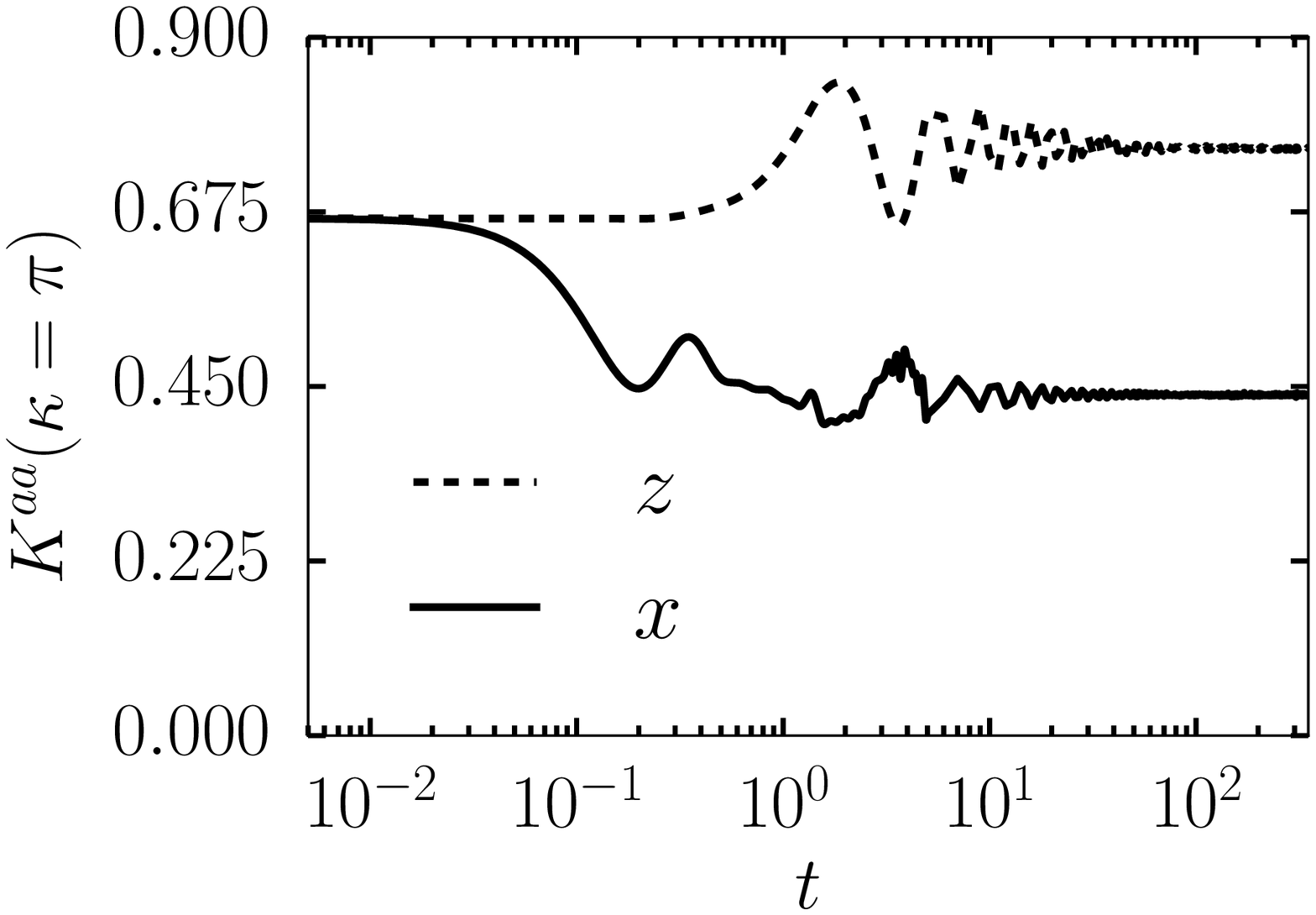}}
	\qquad
	\subfloat[The values of the structure factor in the ground state of the Heisenberg Hamiltonian without {[with]} an additional staggered field is denoted by $K^{aa}_{0}(\kappa)$ {[$K^{aa}_{\mathrm{st}}(\kappa)$]} and indicated by square {[star]} markers. $K^{aa}_{\infty}(\kappa)$ (filled circles) denotes the simulation data evaluated at $t=10^3$.
	]
	{\label{fig:Kaa_comp}\includegraphics[scale=0.3]{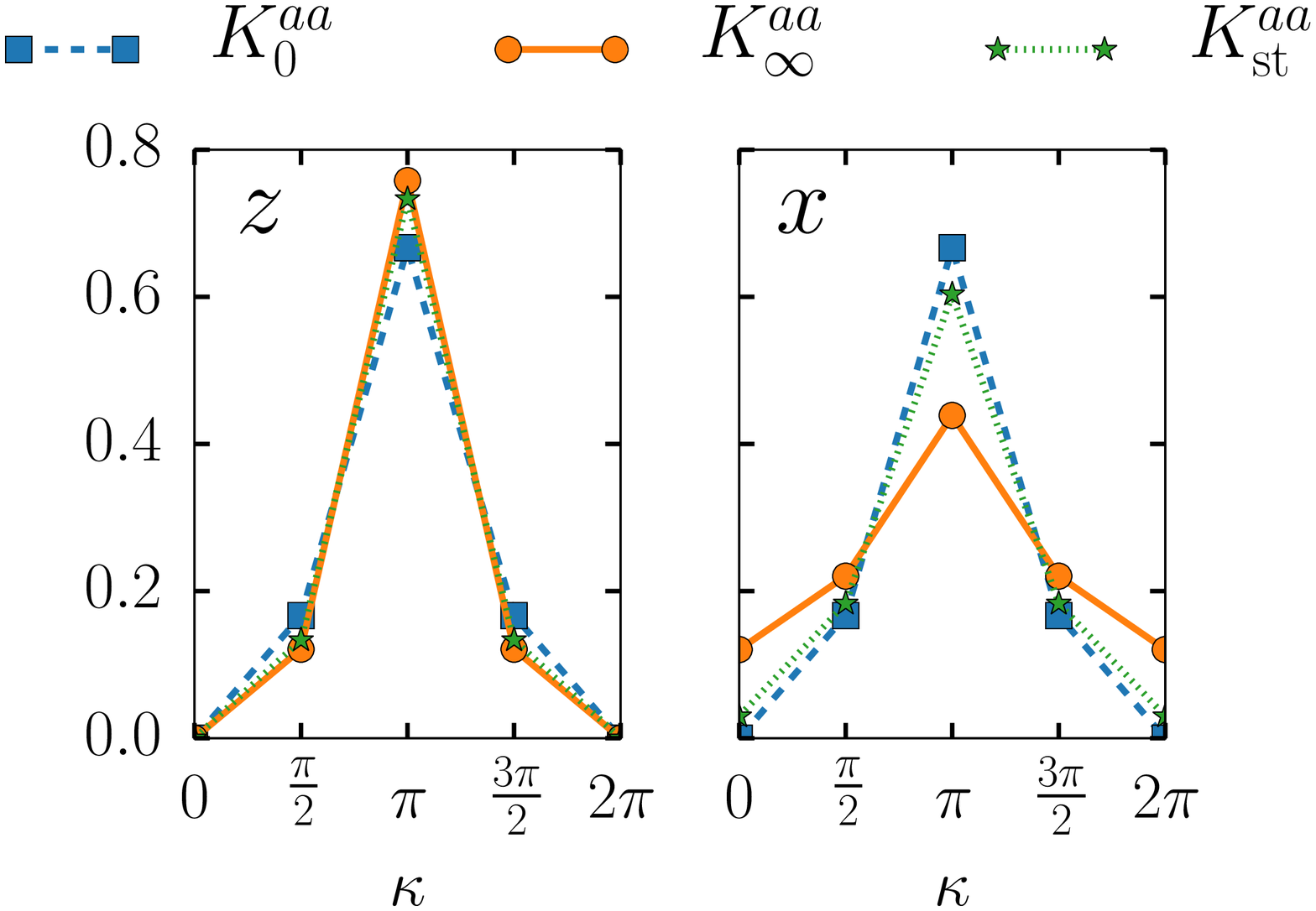}}
	\caption{Simulation data of model $B$: a $N_S=4$ antiferromagnetic spin ring, whereby spin $S_1$ is connected to ${\cal E}_1$ containing $N_{{\cal E}_1}=6$ spins, and the entire central system is in contact with ${\cal E}_2$, a $N_{{\cal E}_2}=12$ spin state that resembles a thermal reservoir at $\beta=50/J_S$. 
${\cal E}_1$ (${\cal E}_2$) is Ising coupled to the central system with random uniform (random binary) strength $I=20J_S$ ($I^\prime=0.1J_S$), and without (with) intra-environment coupling (of strength $K=0.1J_S$).
The	figures show the static structure factor [see Eq.~(\ref{eq:Kaa})], whereby the component $a$ is indicated in the panels.}
\end{figure*}
To exemplify the dynamic process whereby the decoherence of a local spin enhances antiferromagnetic order by generating a de facto staggered field, consider now model $B$ (Sec.~\ref{sec:model_B}).
To reiterate, a low-energy description in terms of the Lieb-Mattis model can not be expected to hold for $d=1$ dimensions, not even approximately. However, by writing $\bm{S}_A = \bm{S}_1 + \bm{S}_3$ and $\bm{S}_B = \bm{S}_2 + \bm{S}_4$ shows that the $N_S=4$ Heisenberg ring (of model $B$) is special and that it coincides with the Lieb-Mattis model exactly.

\subsection{Results model $B$}
Strictly speaking, the spontaneous breaking of symmetry can only occur in the thermodynamic limit. 
The results will therefore be compared to an equivalent system that includes a staggered field $H_S^\prime = H_S^{\bigcirc}  +  J_S/4 \cdot M^z$. 
To quantify the degree of magnetic order, the static structure factor~\cite{ZHU05}
\begin{equation}\label{eq:Kaa}
K^{ab}(\bm{\kappa}) = \langle S^a(\bm{\kappa}) S^b(-\bm{\kappa}) \rangle \, , 
\end{equation}
is used, whereby $\bm{\kappa}=\pi$ corresponds to the antiferromagnetic reciprocal lattice vector on the bipartite chain.
In Fig.~\ref{fig:Kt} the static structure factor is shown at the magnetic reciprocal lattice. Three different regimes can be identified: (i) $t \sim 10^{-1}$ whereby spin $S_1$ is decohered by ${\cal E}_1$ resulting in a reduction of $K^{xx}(\pi)$, (ii) $t \sim 1$ with the DW dominating the dynamics as evidenced by the oscillations, and finally (iii) $t > 10$ where the entire system decoheres due to ${\cal E}_2$, causing the quantum oscillations to be quenched. 

In Fig.~\ref{fig:Kaa_comp} the magnetic ordering of the decohered system is compared to the GS of $H_S^{\bigcirc} $ and that of $H_S^\prime$. The left panel indicates that the enhancement of antiferromagnetic order along the $z$-axis is slightly higher than in the case of a staggered field. 
On the right panel one finds that the magnetic order along the $x$-direction is significantly reduced in comparison to the ground state of the Hamiltonian with a staggered field, $H_S^\prime$. 
This can be attributed to the small size of the CS, since spin $S_1$---which carries significant weight in Eq.~(\ref{eq:Kaa}) for $N_S=4$ spins---becomes completely uncorrelated along the $x$-axis.
\begin{figure}
	\centering
	\includegraphics[scale=0.3]{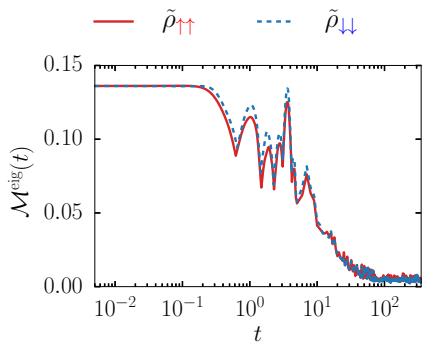}
	\caption{Loss of phase coherence in model $B$. The maximum off diagonal component [see Eq.~(\ref{eq:max_coh})] of the density matrix $\tilde{\rho}_{\uparrow \uparrow}$ ($\tilde{\rho}_{\downarrow \downarrow}$) is evaluated in the basis diagonalising ${H^\prime_{\mathrm{eff}}}^+$ (${H^\prime_{\mathrm{eff}}}^-$).}
	\label{fig:Heff_coh}
\end{figure}
Further support for the claim that the remaining CS is described by the effective Hamiltonian Eq.~(\ref{eq:HLM_eff}) can be obtained by analysing $\tilde{\rho}_{nn}$. If this assumption is correct, then, according to decoherence theory~\cite{PAZ99, SCHL07},  $\tilde{\rho}_{\uparrow \uparrow}$ ($\tilde{\rho}_{\downarrow \downarrow}$) is expected to become diagonal in ${H^\prime_{\mathrm{eff}}}^+$ (${H^\prime_{\mathrm{eff}}}^-$) upon identifying $J^\prime/N=J_S$ and $N = N_S$ \{for a detailed discussion of decoherence in the HH see also Ref.~\cite{DONK17}\}. 
To measure the loss of coherence in the eigen basis, for each time step $t$, the maximum off-diagonal component $|\rho_{n \neq m}(t)|$,
\begin{equation}\label{eq:max_coh}
{\cal M}^{\mathrm{eig}}(t) = \mathrm{max}_{n\neq m} \left[|\langle E_n|\rho(t)|E_m\rangle |\right] \, ,
\end{equation}
is calculated.  Here, the set $\{|E_n\rangle\}$ refer to the eigenstates 
of ${H^\prime_{\mathrm{eff}}}^+$ (${H^\prime_{\mathrm{eff}}}^-$), the effective Hamiltonian of 
$\tilde{\rho}_{\uparrow \uparrow}$ ($\tilde{\rho}_{\downarrow \downarrow}$).
(N.B. coherence is basis dependent; Eqs.~(\ref{eq:coh_loc}) and (\ref{eq:coh_glb}) referred instead to the computational basis.)
The simulation results are shown in Fig.~\ref{fig:Heff_coh}.
Decoherence is indeed observed in Fig.~\ref{fig:Heff_coh}, thereby corroborating the picture in which local decoherence creates an effective staggered field.

Finally, Figs.~\ref{fig:Kt} and \ref{fig:Heff_coh} indicate a decoherence and relaxation timescale of $t \sim 10^2$ (in dimensionless units) for this model. Using an exchange constant $J \sim 6$ meV, as measured in STM experiments~\cite{HIRJ06} (and ignoring for the moment the presence of magnetic anisotropic terms), estimates that antiferromagnetic order develops in $\tau \sim 10$ ps in model $B$.

\section{Discussion and Conclusion}\label{sec:discussion}
Justification of anomalous fields, that single out the classical symmetry-broken states, was often sought in heuristic arguments. For example, in Ref.~\cite{KOMA94} it was suggested that thermal disturbances select states with negligible fluctuation in intensive bulk quantities as the only stable low-energy superpositions.
Here, the possibility of antiferromagnetic order by repeated local measurement~\cite{KATS01} was explored, without the need for a staggered field. 
Within the decoherence framework, continuous---quantum Zeno---measurement was achieved by rather modest environments, containing as little as 7 or 8 spin-1/2 particles, and moderately strong environment coupling ($I=20J_S$ in units of exchange constant $J_S$).
Accordingly, the quantum Zeno picture was applied to a class of isotropic exchange antiferromagnets whereby the low-energy configuration can approximately be described using the Lieb-Mattis (LM) Hamiltonian. To exemplify the enhancement of antiferromagnetic order from the decoherence point of view, the dynamics of a small magnetic structure was analysed whereby a local spin is strongly coupled to an environment.

A decisive parameter that determines whether sublattices can be pinned is the dimensionality $d$. From the linear spin wave perspective the $d \geq 2$ bound follows from the requirement that the energy levels responsible for symmetry breaking are in the thermodynamic limit well separated from excitations that lead to local modulation of the magnetic order (see also the discussion in Sec.~\ref{sec:staggered_field}). Complementary to this, numerical diagonalisation studies of various finite $d=2$ lattices indicates that the approximate picture provided by linear spin wave theory captures the low-energy behaviour of the system surprisingly well~\cite{LHUI05}. In these cases, the analysis leading to Eq.~(\ref{eq:HLM_eff}) seems justified, provided that the static structure factor of the system does not vanish~\cite{BERN92}. In the $d=1$ Heisenberg chain on the contrary, whereby the ground state spin correlations decay algebraically~\cite{BOGO86,PARK10}, repeated measurements are unable to pin down a sublattice~\cite{DONK16}. Only after reducing the quantum fluctuations---by introducing, e.g., anisotropic coupling as done in Ref.~\cite{DONK16}---is one able to create quasi-stable sublattices from a measurement~\cite{DONK16}. 
In this study, magnetic ordering resulted from the decoherence of a small LM magnet, but it is important to note that the staggered magnetisation in the LM model is like a classical vector with zero fluctuation~\cite{KAPL89}.

Our exposition is admittedly somewhat artificial from an experimental point of view. One might argue that a system-environment coupling strength of $I=20J_S$ is unphysically large. But the relevant parameter is the decoherence time scale $\tau$ (as discussed in Sec.~\ref{sec:local_DW}) which depends on the interaction strength $I$, the size of the environment, and possibly other parameters. Thus, $I$ can be small if the environment is sufficiently large.
Secondly, realistic systems usually contain spatially localised impurities or magnetic isotopes that continuously monitor (parts of) the system (see for example Ref.~\cite{HANS08} for the analysis of nitrogen-vacancy centres in diamond). In this sense, the continuous local measurement strategy is not entirely unrealistic. 
It is hoped that this work will pave the way towards more realistic descriptions of local decoherence in antiferromagnets.

\section*{Acknowledgements}
MIK and HCD acknowledge financial support by the European Research Council, project 338957 FEMTO/NANO.

\bibliography{refs}

\begin{thebibliography}{77}%
\makeatletter
\providecommand \@ifxundefined [1]{%
 \@ifx{#1\undefined}
}%
\providecommand \@ifnum [1]{%
 \ifnum #1\expandafter \@firstoftwo
 \else \expandafter \@secondoftwo
 \fi
}%
\providecommand \@ifx [1]{%
 \ifx #1\expandafter \@firstoftwo
 \else \expandafter \@secondoftwo
 \fi
}%
\providecommand \natexlab [1]{#1}%
\providecommand \enquote  [1]{``#1''}%
\providecommand \bibnamefont  [1]{#1}%
\providecommand \bibfnamefont [1]{#1}%
\providecommand \citenamefont [1]{#1}%
\providecommand \href@noop [0]{\@secondoftwo}%
\providecommand \href [0]{\begingroup \@sanitize@url \@href}%
\providecommand \@href[1]{\@@startlink{#1}\@@href}%
\providecommand \@@href[1]{\endgroup#1\@@endlink}%
\providecommand \@sanitize@url [0]{\catcode `\\12\catcode `\$12\catcode
  `\&12\catcode `\#12\catcode `\^12\catcode `\_12\catcode `\%12\relax}%
\providecommand \@@startlink[1]{}%
\providecommand \@@endlink[0]{}%
\providecommand \url  [0]{\begingroup\@sanitize@url \@url }%
\providecommand \@url [1]{\endgroup\@href {#1}{\urlprefix }}%
\providecommand \urlprefix  [0]{URL }%
\providecommand \Eprint [0]{\href }%
\providecommand \doibase [0]{http://dx.doi.org/}%
\providecommand \selectlanguage [0]{\@gobble}%
\providecommand \bibinfo  [0]{\@secondoftwo}%
\providecommand \bibfield  [0]{\@secondoftwo}%
\providecommand \translation [1]{[#1]}%
\providecommand \BibitemOpen [0]{}%
\providecommand \bibitemStop [0]{}%
\providecommand \bibitemNoStop [0]{.\EOS\space}%
\providecommand \EOS [0]{\spacefactor3000\relax}%
\providecommand \BibitemShut  [1]{\csname bibitem#1\endcsname}%
\let\auto@bib@innerbib\@empty
\bibitem [{\citenamefont {Joos}\ \emph {et~al.}(2003)\citenamefont {Joos},
  \citenamefont {Zeh}, \citenamefont {Kiefer}, \citenamefont {Giulini},
  \citenamefont {Kupsch},\ and\ \citenamefont {Stamatescu}}]{JOOS03}%
  \BibitemOpen
  \bibfield  {author} {\bibinfo {author} {\bibfnamefont {E.}~\bibnamefont
  {Joos}}, \bibinfo {author} {\bibfnamefont {H.~D.}\ \bibnamefont {Zeh}},
  \bibinfo {author} {\bibfnamefont {C.}~\bibnamefont {Kiefer}}, \bibinfo
  {author} {\bibfnamefont {D.~J.~W.}\ \bibnamefont {Giulini}}, \bibinfo
  {author} {\bibfnamefont {J.}~\bibnamefont {Kupsch}}, \ and\ \bibinfo {author}
  {\bibfnamefont {I.~O.}\ \bibnamefont {Stamatescu}},\ }\href@noop {} {\emph
  {\bibinfo {title} {Decoherence and the Appearance of a Classical World in
  Quantum Theory}}},\ \bibinfo {edition} {2nd}\ ed.\ (\bibinfo  {publisher}
  {Springer Berlin Heidelberg},\ \bibinfo {year} {2003})\BibitemShut {NoStop}%
\bibitem [{\citenamefont {Zurek}(2003)}]{ZURE03}%
  \BibitemOpen
  \bibfield  {author} {\bibinfo {author} {\bibfnamefont {W.~H.}\ \bibnamefont
  {Zurek}},\ }\href {\doibase 10.1103/RevModPhys.75.715} {\bibfield  {journal}
  {\bibinfo  {journal} {Rev. Mod. Phys.}\ }\textbf {\bibinfo {volume} {75}},\
  \bibinfo {pages} {715} (\bibinfo {year} {2003})}\BibitemShut {NoStop}%
\bibitem [{\citenamefont {Schlosshauer}(2007)}]{SCHL07}%
  \BibitemOpen
  \bibfield  {author} {\bibinfo {author} {\bibfnamefont {M.~A.}\ \bibnamefont
  {Schlosshauer}},\ }\href@noop {} {\emph {\bibinfo {title} {Decoherence: and
  the quantum-to-classical transition}}}\ (\bibinfo  {publisher} {Springer},\
  \bibinfo {year} {2007})\BibitemShut {NoStop}%
\bibitem [{\citenamefont {D{\"u}r}\ and\ \citenamefont
  {Briegel}(2004)}]{DUR04}%
  \BibitemOpen
  \bibfield  {author} {\bibinfo {author} {\bibfnamefont {W.}~\bibnamefont
  {D{\"u}r}}\ and\ \bibinfo {author} {\bibfnamefont {H.-J.}\ \bibnamefont
  {Briegel}},\ }\href {\doibase 10.1103/PhysRevLett.92.180403} {\bibfield
  {journal} {\bibinfo  {journal} {Phys. Rev. Lett.}\ }\textbf {\bibinfo
  {volume} {92}},\ \bibinfo {pages} {180403} (\bibinfo {year}
  {2004})}\BibitemShut {NoStop}%
\bibitem [{\citenamefont {Carvalho}\ \emph {et~al.}(2004)\citenamefont
  {Carvalho}, \citenamefont {Mintert},\ and\ \citenamefont
  {Buchleitner}}]{CARV04}%
  \BibitemOpen
  \bibfield  {author} {\bibinfo {author} {\bibfnamefont {A.~R.~R.}\
  \bibnamefont {Carvalho}}, \bibinfo {author} {\bibfnamefont {F.}~\bibnamefont
  {Mintert}}, \ and\ \bibinfo {author} {\bibfnamefont {A.}~\bibnamefont
  {Buchleitner}},\ }\href {\doibase 10.1103/PhysRevLett.93.230501} {\bibfield
  {journal} {\bibinfo  {journal} {Phys. Rev. Lett.}\ }\textbf {\bibinfo
  {volume} {93}},\ \bibinfo {pages} {230501} (\bibinfo {year}
  {2004})}\BibitemShut {NoStop}%
\bibitem [{\citenamefont {Yu}\ and\ \citenamefont {Eberly}(2009)}]{YU09}%
  \BibitemOpen
  \bibfield  {author} {\bibinfo {author} {\bibfnamefont {T.}~\bibnamefont
  {Yu}}\ and\ \bibinfo {author} {\bibfnamefont {J.~H.}\ \bibnamefont
  {Eberly}},\ }\href {\doibase 10.1126/science.1167343} {\bibfield  {journal}
  {\bibinfo  {journal} {Science}\ }\textbf {\bibinfo {volume} {323}},\ \bibinfo
  {pages} {598} (\bibinfo {year} {2009})}\BibitemShut {NoStop}%
\bibitem [{\citenamefont {Borras}\ \emph {et~al.}(2009)\citenamefont {Borras},
  \citenamefont {Majtey}, \citenamefont {Plastino}, \citenamefont {Casas},\
  and\ \citenamefont {Plastino}}]{BORR09}%
  \BibitemOpen
  \bibfield  {author} {\bibinfo {author} {\bibfnamefont {A.}~\bibnamefont
  {Borras}}, \bibinfo {author} {\bibfnamefont {A.~P.}\ \bibnamefont {Majtey}},
  \bibinfo {author} {\bibfnamefont {A.~R.}\ \bibnamefont {Plastino}}, \bibinfo
  {author} {\bibfnamefont {M.}~\bibnamefont {Casas}}, \ and\ \bibinfo {author}
  {\bibfnamefont {A.}~\bibnamefont {Plastino}},\ }\href {\doibase
  10.1103/PhysRevA.79.022108} {\bibfield  {journal} {\bibinfo  {journal} {Phys.
  Rev. A}\ }\textbf {\bibinfo {volume} {79}},\ \bibinfo {pages} {022108}
  (\bibinfo {year} {2009})}\BibitemShut {NoStop}%
\bibitem [{\citenamefont {Aolita}\ \emph {et~al.}(2015)\citenamefont {Aolita},
  \citenamefont {De~Melo},\ and\ \citenamefont {Davidovich}}]{AOLI15}%
  \BibitemOpen
  \bibfield  {author} {\bibinfo {author} {\bibfnamefont {L.}~\bibnamefont
  {Aolita}}, \bibinfo {author} {\bibfnamefont {F.}~\bibnamefont {De~Melo}}, \
  and\ \bibinfo {author} {\bibfnamefont {L.}~\bibnamefont {Davidovich}},\
  }\href {\doibase 10.1088/0034-4885/78/4/042001} {\bibfield  {journal}
  {\bibinfo  {journal} {Rep. Prog. Phys.}\ }\textbf {\bibinfo {volume} {78}},\
  \bibinfo {pages} {042001} (\bibinfo {year} {2015})}\BibitemShut {NoStop}%
\bibitem [{\citenamefont {Prokof'Ev}\ and\ \citenamefont
  {Stamp}(1996)}]{PROK96}%
  \BibitemOpen
  \bibfield  {author} {\bibinfo {author} {\bibfnamefont {N.~V.}\ \bibnamefont
  {Prokof'Ev}}\ and\ \bibinfo {author} {\bibfnamefont {P.~C.~E.}\ \bibnamefont
  {Stamp}},\ }\href {\doibase 10.1007/BF00754094} {\bibfield  {journal}
  {\bibinfo  {journal} {J. Low Temp. Phys.}\ }\textbf {\bibinfo {volume}
  {104}},\ \bibinfo {pages} {143} (\bibinfo {year} {1996})}\BibitemShut
  {NoStop}%
\bibitem [{\citenamefont {Prokof'ev}\ and\ \citenamefont
  {Stamp}(2000)}]{PROK00}%
  \BibitemOpen
  \bibfield  {author} {\bibinfo {author} {\bibfnamefont {N.~V.}\ \bibnamefont
  {Prokof'ev}}\ and\ \bibinfo {author} {\bibfnamefont {P.~C.~E.}\ \bibnamefont
  {Stamp}},\ }\href {\doibase 10.1088/0034-4885/63/4/204} {\bibfield  {journal}
  {\bibinfo  {journal} {Rep. Prog. Phys.}\ }\textbf {\bibinfo {volume} {63}},\
  \bibinfo {pages} {669} (\bibinfo {year} {2000})}\BibitemShut {NoStop}%
\bibitem [{\citenamefont {Gauyacq}\ and\ \citenamefont
  {Lorente}(2015)}]{GAUY15}%
  \BibitemOpen
  \bibfield  {author} {\bibinfo {author} {\bibfnamefont {J.-P.}\ \bibnamefont
  {Gauyacq}}\ and\ \bibinfo {author} {\bibfnamefont {N.}~\bibnamefont
  {Lorente}},\ }\href {\doibase 10.1088/0953-8984/27/45/455301} {\bibfield
  {journal} {\bibinfo  {journal} {J. Phys.: Condens. Matter}\ }\textbf
  {\bibinfo {volume} {27}},\ \bibinfo {pages} {455301} (\bibinfo {year}
  {2015})}\BibitemShut {NoStop}%
\bibitem [{\citenamefont {Delgado}\ \emph {et~al.}(2015)\citenamefont
  {Delgado}, \citenamefont {Loth}, \citenamefont {Zielinski},\ and\
  \citenamefont {Fern{\'a}ndez-Rossier}}]{DELG15}%
  \BibitemOpen
  \bibfield  {author} {\bibinfo {author} {\bibfnamefont {F.}~\bibnamefont
  {Delgado}}, \bibinfo {author} {\bibfnamefont {S.}~\bibnamefont {Loth}},
  \bibinfo {author} {\bibfnamefont {M.}~\bibnamefont {Zielinski}}, \ and\
  \bibinfo {author} {\bibfnamefont {J.}~\bibnamefont {Fern{\'a}ndez-Rossier}},\
  }\href {\doibase 10.1209/0295-5075/109/57001} {\bibfield  {journal} {\bibinfo
   {journal} {Europhys. Lett.}\ }\textbf {\bibinfo {volume} {109}},\ \bibinfo
  {pages} {57001} (\bibinfo {year} {2015})}\BibitemShut {NoStop}%
\bibitem [{\citenamefont {Delgado}\ and\ \citenamefont
  {Fern{\'a}ndez-Rossier}(2017)}]{DELG17}%
  \BibitemOpen
  \bibfield  {author} {\bibinfo {author} {\bibfnamefont {F.}~\bibnamefont
  {Delgado}}\ and\ \bibinfo {author} {\bibfnamefont {J.}~\bibnamefont
  {Fern{\'a}ndez-Rossier}},\ }\href {\doibase 10.1016/j.progsurf.2016.12.001}
  {\bibfield  {journal} {\bibinfo  {journal} {Prog. Surf. Sci.}\ }\textbf
  {\bibinfo {volume} {92}},\ \bibinfo {pages} {40 } (\bibinfo {year}
  {2017})}\BibitemShut {NoStop}%
\bibitem [{\citenamefont {Leggett}\ \emph {et~al.}(1987)\citenamefont
  {Leggett}, \citenamefont {Chakravarty}, \citenamefont {Dorsey}, \citenamefont
  {Fisher}, \citenamefont {Garg},\ and\ \citenamefont {Zwerger}}]{LEGG87}%
  \BibitemOpen
  \bibfield  {author} {\bibinfo {author} {\bibfnamefont {A.~J.}\ \bibnamefont
  {Leggett}}, \bibinfo {author} {\bibfnamefont {S.}~\bibnamefont
  {Chakravarty}}, \bibinfo {author} {\bibfnamefont {A.~T.}\ \bibnamefont
  {Dorsey}}, \bibinfo {author} {\bibfnamefont {M.~P.~A.}\ \bibnamefont
  {Fisher}}, \bibinfo {author} {\bibfnamefont {A.}~\bibnamefont {Garg}}, \ and\
  \bibinfo {author} {\bibfnamefont {W.}~\bibnamefont {Zwerger}},\ }\href@noop
  {} {\bibfield  {journal} {\bibinfo  {journal} {Rev. Mod. Phys.}\ }\textbf
  {\bibinfo {volume} {59}},\ \bibinfo {pages} {1} (\bibinfo {year}
  {1987})}\BibitemShut {NoStop}%
\bibitem [{\citenamefont {Zhang}\ \emph {et~al.}(2007)\citenamefont {Zhang},
  \citenamefont {Konstantinidis}, \citenamefont {Al-Hassanieh},\ and\
  \citenamefont {Dobrovitski}}]{ZHAN07}%
  \BibitemOpen
  \bibfield  {author} {\bibinfo {author} {\bibfnamefont {W.}~\bibnamefont
  {Zhang}}, \bibinfo {author} {\bibfnamefont {N.}~\bibnamefont
  {Konstantinidis}}, \bibinfo {author} {\bibfnamefont {K.~A.}\ \bibnamefont
  {Al-Hassanieh}}, \ and\ \bibinfo {author} {\bibfnamefont {V.~V.}\
  \bibnamefont {Dobrovitski}},\ }\href {\doibase 10.1088/0953-8984/19/8/083202}
  {\bibfield  {journal} {\bibinfo  {journal} {J. Phys.: Condens. Matter}\
  }\textbf {\bibinfo {volume} {19}},\ \bibinfo {pages} {083202} (\bibinfo
  {year} {2007})}\BibitemShut {NoStop}%
\bibitem [{\citenamefont {Anderson}(1972)}]{ANDE72}%
  \BibitemOpen
  \bibfield  {author} {\bibinfo {author} {\bibfnamefont {P.~W.}\ \bibnamefont
  {Anderson}},\ }\href {\doibase 10.1126/science.177.4047.393} {\bibfield
  {journal} {\bibinfo  {journal} {Science}\ }\textbf {\bibinfo {volume}
  {177}},\ \bibinfo {pages} {393} (\bibinfo {year} {1972})}\BibitemShut
  {NoStop}%
\bibitem [{\citenamefont {Laughlin}\ and\ \citenamefont
  {Pines}(2000)}]{LAUG00}%
  \BibitemOpen
  \bibfield  {author} {\bibinfo {author} {\bibfnamefont {R.~B.}\ \bibnamefont
  {Laughlin}}\ and\ \bibinfo {author} {\bibfnamefont {D.}~\bibnamefont
  {Pines}},\ }\href@noop {} {\bibfield  {journal} {\bibinfo  {journal} {Proc.
  Natl. Acad. Sci. U.S.A.}\ }\textbf {\bibinfo {volume} {97}},\ \bibinfo
  {pages} {28} (\bibinfo {year} {2000})}\BibitemShut {NoStop}%
\bibitem [{\citenamefont {Laughlin}(2006)}]{LAUG06}%
  \BibitemOpen
  \bibfield  {author} {\bibinfo {author} {\bibfnamefont {R.~B.}\ \bibnamefont
  {Laughlin}},\ }\href@noop {} {\emph {\bibinfo {title} {A Different Universe:
  Reinventing Physics From the Bottom Down}}}\ (\bibinfo  {publisher} {Basic
  Books},\ \bibinfo {year} {2006})\BibitemShut {NoStop}%
\bibitem [{\citenamefont {Anderson}(1964)}]{ANDE64}%
  \BibitemOpen
  \bibfield  {author} {\bibinfo {author} {\bibfnamefont {P.~W.}\ \bibnamefont
  {Anderson}},\ }\href@noop {} {\emph {\bibinfo {title} {Concepts in Solids:
  Lectures on the Theory of Solids}}},\ Frontiers in Physics : a lecture note
  and reprint series\ (\bibinfo  {publisher} {W.A. Benjamin},\ \bibinfo {year}
  {1964})\BibitemShut {NoStop}%
\bibitem [{\citenamefont {Koma}\ and\ \citenamefont {Tasaki}(1994)}]{KOMA94}%
  \BibitemOpen
  \bibfield  {author} {\bibinfo {author} {\bibfnamefont {T.}~\bibnamefont
  {Koma}}\ and\ \bibinfo {author} {\bibfnamefont {H.}~\bibnamefont {Tasaki}},\
  }\href {\doibase 10.1007/BF02188685} {\bibfield  {journal} {\bibinfo
  {journal} {J. Stat. Phys.}\ }\textbf {\bibinfo {volume} {76}},\ \bibinfo
  {pages} {745} (\bibinfo {year} {1994})}\BibitemShut {NoStop}%
\bibitem [{\citenamefont {Vonsovsky}\ and\ \citenamefont
  {Svirsky}(1969)}]{VONS69}%
  \BibitemOpen
  \bibfield  {author} {\bibinfo {author} {\bibfnamefont {S.~V.}\ \bibnamefont
  {Vonsovsky}}\ and\ \bibinfo {author} {\bibfnamefont {M.~S.}\ \bibnamefont
  {Svirsky}},\ }\href@noop {} {\bibfield  {journal} {\bibinfo  {journal} {Zh.
  \'Eksp. Teor. Fiz.}\ }\textbf {\bibinfo {volume} {57}},\ \bibinfo {pages}
  {251} (\bibinfo {year} {1969})}\BibitemShut {NoStop}%
\bibitem [{\citenamefont {Vonsovsky}(1974)}]{VONS74}%
  \BibitemOpen
  \bibfield  {author} {\bibinfo {author} {\bibfnamefont {S.~V.}\ \bibnamefont
  {Vonsovsky}},\ }\href@noop {} {\emph {\bibinfo {title} {Magnetism}}},\
  Vol.~\bibinfo {volume} {2}\ (\bibinfo  {publisher} {Wiley},\ \bibinfo {year}
  {1974})\BibitemShut {NoStop}%
\bibitem [{\citenamefont {Irkhin}\ and\ \citenamefont
  {Katsnelson}(1986)}]{IRKH86}%
  \BibitemOpen
  \bibfield  {author} {\bibinfo {author} {\bibfnamefont {V.~Y.}\ \bibnamefont
  {Irkhin}}\ and\ \bibinfo {author} {\bibfnamefont {M.~I.}\ \bibnamefont
  {Katsnelson}},\ }\href {\doibase 10.1007/BF01323431} {\bibfield  {journal}
  {\bibinfo  {journal} {Z. Phys. B}\ }\textbf {\bibinfo {volume} {62}},\
  \bibinfo {pages} {201} (\bibinfo {year} {1986})}\BibitemShut {NoStop}%
\bibitem [{\citenamefont {Lieb}\ and\ \citenamefont {Mattis}(1962)}]{LIEB62}%
  \BibitemOpen
  \bibfield  {author} {\bibinfo {author} {\bibfnamefont {E.}~\bibnamefont
  {Lieb}}\ and\ \bibinfo {author} {\bibfnamefont {D.}~\bibnamefont {Mattis}},\
  }\href {\doibase 10.1063/1.1724276} {\bibfield  {journal} {\bibinfo
  {journal} {J. Math. Phys.}\ }\textbf {\bibinfo {volume} {3}},\ \bibinfo
  {pages} {749} (\bibinfo {year} {1962})}\BibitemShut {NoStop}%
\bibitem [{\citenamefont {Jones}(1998)}]{JONE98}%
  \BibitemOpen
  \bibfield  {author} {\bibinfo {author} {\bibfnamefont {H.}~\bibnamefont
  {Jones}},\ }\href@noop {} {\emph {\bibinfo {title} {Groups, Representations
  and Physics}}}\ (\bibinfo  {publisher} {Taylor \& Fracis Group},\ \bibinfo
  {year} {1998})\BibitemShut {NoStop}%
\bibitem [{\citenamefont {Bernu}\ \emph {et~al.}(1992)\citenamefont {Bernu},
  \citenamefont {Lhuillier},\ and\ \citenamefont {Pierre}}]{BERN92}%
  \BibitemOpen
  \bibfield  {author} {\bibinfo {author} {\bibfnamefont {B.}~\bibnamefont
  {Bernu}}, \bibinfo {author} {\bibfnamefont {C.}~\bibnamefont {Lhuillier}}, \
  and\ \bibinfo {author} {\bibfnamefont {L.}~\bibnamefont {Pierre}},\ }\href
  {\doibase 10.1103/PhysRevLett.69.2590} {\bibfield  {journal} {\bibinfo
  {journal} {Phys. Rev. Lett.}\ }\textbf {\bibinfo {volume} {69}},\ \bibinfo
  {pages} {2590} (\bibinfo {year} {1992})}\BibitemShut {NoStop}%
\bibitem [{\citenamefont {Anderson}(1952)}]{ANDE52}%
  \BibitemOpen
  \bibfield  {author} {\bibinfo {author} {\bibfnamefont {P.~W.}\ \bibnamefont
  {Anderson}},\ }\href {\doibase 10.1103/PhysRev.86.694} {\bibfield  {journal}
  {\bibinfo  {journal} {Phys. Rev.}\ }\textbf {\bibinfo {volume} {86}},\
  \bibinfo {pages} {694} (\bibinfo {year} {1952})}\BibitemShut {NoStop}%
\bibitem [{\citenamefont {Anderson}(1951)}]{ANDE51}%
  \BibitemOpen
  \bibfield  {author} {\bibinfo {author} {\bibfnamefont {P.~W.}\ \bibnamefont
  {Anderson}},\ }\href {\doibase 10.1103/PhysRev.83.1260} {\bibfield  {journal}
  {\bibinfo  {journal} {Phys. Rev.}\ }\textbf {\bibinfo {volume} {83}},\
  \bibinfo {pages} {1260} (\bibinfo {year} {1951})}\BibitemShut {NoStop}%
\bibitem [{\citenamefont {Bogolubov}(1960)}]{BOGO60}%
  \BibitemOpen
  \bibfield  {author} {\bibinfo {author} {\bibfnamefont {N.~N.}\ \bibnamefont
  {Bogolubov}},\ }\href {\doibase 10.1016/0031-8914(60)90196-8} {\bibfield
  {journal} {\bibinfo  {journal} {Physica}\ }\textbf {\bibinfo {volume} {26}},\
  \bibinfo {pages} {S1} (\bibinfo {year} {1960})}\BibitemShut {NoStop}%
\bibitem [{\citenamefont {Oshikawa}\ and\ \citenamefont
  {Affleck}(1997)}]{OSHI97}%
  \BibitemOpen
  \bibfield  {author} {\bibinfo {author} {\bibfnamefont {M.}~\bibnamefont
  {Oshikawa}}\ and\ \bibinfo {author} {\bibfnamefont {I.}~\bibnamefont
  {Affleck}},\ }\href {\doibase 10.1103/PhysRevLett.79.2883} {\bibfield
  {journal} {\bibinfo  {journal} {Phys. Rev. Lett.}\ }\textbf {\bibinfo
  {volume} {79}},\ \bibinfo {pages} {2883} (\bibinfo {year}
  {1997})}\BibitemShut {NoStop}%
\bibitem [{\citenamefont {Ziman}(1952)}]{ZIMA52}%
  \BibitemOpen
  \bibfield  {author} {\bibinfo {author} {\bibfnamefont {J.~M.}\ \bibnamefont
  {Ziman}},\ }\href {\doibase 10.1088/0370-1298/65/7/309} {\bibfield  {journal}
  {\bibinfo  {journal} {Proc. Phys. Soc.}\ }\textbf {\bibinfo {volume} {65}},\
  \bibinfo {pages} {540} (\bibinfo {year} {1952})}\BibitemShut {NoStop}%
\bibitem [{\citenamefont {Kuzemsky}(2010)}]{KUZE10}%
  \BibitemOpen
  \bibfield  {author} {\bibinfo {author} {\bibfnamefont {A.~L.}\ \bibnamefont
  {Kuzemsky}},\ }\href {\doibase 10.1142/S0217979210055378} {\bibfield
  {journal} {\bibinfo  {journal} {Int. J. Mod. Phys. B}\ }\textbf {\bibinfo
  {volume} {24}},\ \bibinfo {pages} {835} (\bibinfo {year} {2010})}\BibitemShut
  {NoStop}%
\bibitem [{\citenamefont {Huang}(2000)}]{HUAN00}%
  \BibitemOpen
  \bibfield  {author} {\bibinfo {author} {\bibfnamefont {K.}~\bibnamefont
  {Huang}},\ }\href@noop {} {\emph {\bibinfo {title} {Statistical Mechanics}}}\
  (\bibinfo  {publisher} {John Wiley and Sons},\ \bibinfo {year}
  {2000})\BibitemShut {NoStop}%
\bibitem [{\citenamefont {Pratt}(1961)}]{PRAT61}%
  \BibitemOpen
  \bibfield  {author} {\bibinfo {author} {\bibfnamefont {G.~W.}\ \bibnamefont
  {Pratt}},\ }\href {\doibase 10.1103/PhysRev.122.489} {\bibfield  {journal}
  {\bibinfo  {journal} {Phys. Rev.}\ }\textbf {\bibinfo {volume} {122}},\
  \bibinfo {pages} {489} (\bibinfo {year} {1961})}\BibitemShut {NoStop}%
\bibitem [{\citenamefont {{Lhuillier}}(2005)}]{LHUI05}%
  \BibitemOpen
  \bibfield  {author} {\bibinfo {author} {\bibfnamefont {C.}~\bibnamefont
  {{Lhuillier}}},\ }\href@noop {} {\bibfield  {journal} {\bibinfo  {journal}
  {eprint arXiv:cond-mat/0502464}\ } (\bibinfo {year} {2005})}\BibitemShut
  {NoStop}%
\bibitem [{\citenamefont {Kaiser}\ and\ \citenamefont
  {Peschel}(1989)}]{KAIS89}%
  \BibitemOpen
  \bibfield  {author} {\bibinfo {author} {\bibfnamefont {C.}~\bibnamefont
  {Kaiser}}\ and\ \bibinfo {author} {\bibfnamefont {I.}~\bibnamefont
  {Peschel}},\ }\href {\doibase 10.1088/0305-4470/22/19/018} {\bibfield
  {journal} {\bibinfo  {journal} {J. Phys. A: Math. Gen.}\ }\textbf {\bibinfo
  {volume} {22}},\ \bibinfo {pages} {4257} (\bibinfo {year}
  {1989})}\BibitemShut {NoStop}%
\bibitem [{\citenamefont {Kaplan}\ \emph {et~al.}(1990)\citenamefont {Kaplan},
  \citenamefont {von~der Linden},\ and\ \citenamefont {Horsch}}]{KAPL90}%
  \BibitemOpen
  \bibfield  {author} {\bibinfo {author} {\bibfnamefont {T.~A.}\ \bibnamefont
  {Kaplan}}, \bibinfo {author} {\bibfnamefont {W.}~\bibnamefont {von~der
  Linden}}, \ and\ \bibinfo {author} {\bibfnamefont {P.}~\bibnamefont
  {Horsch}},\ }\href {\doibase 10.1103/PhysRevB.42.4663} {\bibfield  {journal}
  {\bibinfo  {journal} {Phys. Rev. B}\ }\textbf {\bibinfo {volume} {42}},\
  \bibinfo {pages} {4663} (\bibinfo {year} {1990})}\BibitemShut {NoStop}%
\bibitem [{\citenamefont {Dyson}\ \emph {et~al.}(1978)\citenamefont {Dyson},
  \citenamefont {Lieb},\ and\ \citenamefont {Simon}}]{DYSO78}%
  \BibitemOpen
  \bibfield  {author} {\bibinfo {author} {\bibfnamefont {F.~J.}\ \bibnamefont
  {Dyson}}, \bibinfo {author} {\bibfnamefont {E.~H.}\ \bibnamefont {Lieb}}, \
  and\ \bibinfo {author} {\bibfnamefont {B.}~\bibnamefont {Simon}},\ }\href
  {\doibase 10.1007/BF01106729} {\bibfield  {journal} {\bibinfo  {journal} {J.
  Stat. Phys.}\ }\textbf {\bibinfo {volume} {18}},\ \bibinfo {pages} {335}
  (\bibinfo {year} {1978})}\BibitemShut {NoStop}%
\bibitem [{\citenamefont {Kennedy}\ \emph {et~al.}(1988)\citenamefont
  {Kennedy}, \citenamefont {Lieb},\ and\ \citenamefont {Shastry}}]{KENN88}%
  \BibitemOpen
  \bibfield  {author} {\bibinfo {author} {\bibfnamefont {T.}~\bibnamefont
  {Kennedy}}, \bibinfo {author} {\bibfnamefont {E.~H.}\ \bibnamefont {Lieb}}, \
  and\ \bibinfo {author} {\bibfnamefont {B.~S.}\ \bibnamefont {Shastry}},\
  }\href@noop {} {\bibfield  {journal} {\bibinfo  {journal} {J. Stat. Phys.}\
  }\textbf {\bibinfo {volume} {53}},\ \bibinfo {pages} {1019} (\bibinfo {year}
  {1988})}\BibitemShut {NoStop}%
\bibitem [{\citenamefont {L{\"o}w}(2007)}]{LOW07}%
  \BibitemOpen
  \bibfield  {author} {\bibinfo {author} {\bibfnamefont {U.}~\bibnamefont
  {L{\"o}w}},\ }\href {\doibase 10.1103/PhysRevB.76.220409} {\bibfield
  {journal} {\bibinfo  {journal} {Phys. Rev. B}\ }\textbf {\bibinfo {volume}
  {76}},\ \bibinfo {pages} {220409} (\bibinfo {year} {2007})}\BibitemShut
  {NoStop}%
\bibitem [{\citenamefont {White}\ and\ \citenamefont
  {Chernyshev}(2007)}]{WHIT07}%
  \BibitemOpen
  \bibfield  {author} {\bibinfo {author} {\bibfnamefont {S.~R.}\ \bibnamefont
  {White}}\ and\ \bibinfo {author} {\bibfnamefont {A.~L.}\ \bibnamefont
  {Chernyshev}},\ }\href {\doibase 10.1103/PhysRevLett.99.127004} {\bibfield
  {journal} {\bibinfo  {journal} {Phys. Rev. Lett.}\ }\textbf {\bibinfo
  {volume} {99}},\ \bibinfo {pages} {127004} (\bibinfo {year}
  {2007})}\BibitemShut {NoStop}%
\bibitem [{\citenamefont {Jolicoeur}\ \emph {et~al.}(1990)\citenamefont
  {Jolicoeur}, \citenamefont {Dagotto}, \citenamefont {Gagliano},\ and\
  \citenamefont {Bacci}}]{JOLI90}%
  \BibitemOpen
  \bibfield  {author} {\bibinfo {author} {\bibfnamefont {T.}~\bibnamefont
  {Jolicoeur}}, \bibinfo {author} {\bibfnamefont {E.}~\bibnamefont {Dagotto}},
  \bibinfo {author} {\bibfnamefont {E.}~\bibnamefont {Gagliano}}, \ and\
  \bibinfo {author} {\bibfnamefont {S.}~\bibnamefont {Bacci}},\ }\href
  {\doibase 10.1103/PhysRevB.42.4800} {\bibfield  {journal} {\bibinfo
  {journal} {Phys. Rev. B}\ }\textbf {\bibinfo {volume} {42}},\ \bibinfo
  {pages} {4800} (\bibinfo {year} {1990})}\BibitemShut {NoStop}%
\bibitem [{\citenamefont {Bernu}\ \emph {et~al.}(1994)\citenamefont {Bernu},
  \citenamefont {Lecheminant}, \citenamefont {Lhuillier},\ and\ \citenamefont
  {Pierre}}]{BERN94}%
  \BibitemOpen
  \bibfield  {author} {\bibinfo {author} {\bibfnamefont {B.}~\bibnamefont
  {Bernu}}, \bibinfo {author} {\bibfnamefont {P.}~\bibnamefont {Lecheminant}},
  \bibinfo {author} {\bibfnamefont {C.}~\bibnamefont {Lhuillier}}, \ and\
  \bibinfo {author} {\bibfnamefont {L.}~\bibnamefont {Pierre}},\ }\href
  {\doibase 10.1103/PhysRevB.50.10048} {\bibfield  {journal} {\bibinfo
  {journal} {Phys. Rev. B}\ }\textbf {\bibinfo {volume} {50}},\ \bibinfo
  {pages} {10048} (\bibinfo {year} {1994})}\BibitemShut {NoStop}%
\bibitem [{\citenamefont {Capriotti}\ \emph {et~al.}(1999)\citenamefont
  {Capriotti}, \citenamefont {Trumper},\ and\ \citenamefont
  {Sorella}}]{CAPR99}%
  \BibitemOpen
  \bibfield  {author} {\bibinfo {author} {\bibfnamefont {L.}~\bibnamefont
  {Capriotti}}, \bibinfo {author} {\bibfnamefont {A.~E.}\ \bibnamefont
  {Trumper}}, \ and\ \bibinfo {author} {\bibfnamefont {S.}~\bibnamefont
  {Sorella}},\ }\href {\doibase 10.1103/PhysRevLett.82.3899} {\bibfield
  {journal} {\bibinfo  {journal} {Phys. Rev. Lett.}\ }\textbf {\bibinfo
  {volume} {82}},\ \bibinfo {pages} {3899} (\bibinfo {year}
  {1999})}\BibitemShut {NoStop}%
\bibitem [{\citenamefont {Kaplan}\ \emph {et~al.}(1989)\citenamefont {Kaplan},
  \citenamefont {Horsch},\ and\ \citenamefont {{Von der Linden}}}]{KAPL89}%
  \BibitemOpen
  \bibfield  {author} {\bibinfo {author} {\bibfnamefont {T.~A.}\ \bibnamefont
  {Kaplan}}, \bibinfo {author} {\bibfnamefont {P.}~\bibnamefont {Horsch}}, \
  and\ \bibinfo {author} {\bibfnamefont {W.}~\bibnamefont {{Von der Linden}}},\
  }\href {\doibase 10.1143/JPSJ.58.3894} {\bibfield  {journal} {\bibinfo
  {journal} {J. Phys. Soc. Jpn.}\ }\textbf {\bibinfo {volume} {58}},\ \bibinfo
  {pages} {3894} (\bibinfo {year} {1989})}\BibitemShut {NoStop}%
\bibitem [{\citenamefont {Koma}\ and\ \citenamefont {Tasaki}(1993)}]{KOMA93}%
  \BibitemOpen
  \bibfield  {author} {\bibinfo {author} {\bibfnamefont {T.}~\bibnamefont
  {Koma}}\ and\ \bibinfo {author} {\bibfnamefont {H.}~\bibnamefont {Tasaki}},\
  }\href {\doibase 10.1103/PhysRevLett.70.93} {\bibfield  {journal} {\bibinfo
  {journal} {Phys. Rev. Lett.}\ }\textbf {\bibinfo {volume} {70}},\ \bibinfo
  {pages} {93} (\bibinfo {year} {1993})}\BibitemShut {NoStop}%
\bibitem [{\citenamefont {Astrov}\ \emph {et~al.}(1996)\citenamefont {Astrov},
  \citenamefont {Ermakov}, \citenamefont {Borovik-Romanov}, \citenamefont
  {Kolevatov},\ and\ \citenamefont {Nizhankovskii}}]{ASTR96}%
  \BibitemOpen
  \bibfield  {author} {\bibinfo {author} {\bibfnamefont {D.~N.}\ \bibnamefont
  {Astrov}}, \bibinfo {author} {\bibfnamefont {N.~B.}\ \bibnamefont {Ermakov}},
  \bibinfo {author} {\bibfnamefont {A.~S.}\ \bibnamefont {Borovik-Romanov}},
  \bibinfo {author} {\bibfnamefont {E.~G.}\ \bibnamefont {Kolevatov}}, \ and\
  \bibinfo {author} {\bibfnamefont {V.~I.}\ \bibnamefont {Nizhankovskii}},\
  }\href@noop {} {\bibfield  {journal} {\bibinfo  {journal} {Pis'ma Zh. \'Eksp.
  Teor. Fiz.}\ }\textbf {\bibinfo {volume} {63}},\ \bibinfo {pages} {713}
  (\bibinfo {year} {1996})}\BibitemShut {NoStop}%
\bibitem [{\citenamefont {Loth}\ \emph {et~al.}(2012)\citenamefont {Loth},
  \citenamefont {Baumann}, \citenamefont {Lutz}, \citenamefont {Eigler},\ and\
  \citenamefont {Heinrich}}]{LOTH12}%
  \BibitemOpen
  \bibfield  {author} {\bibinfo {author} {\bibfnamefont {S.}~\bibnamefont
  {Loth}}, \bibinfo {author} {\bibfnamefont {S.}~\bibnamefont {Baumann}},
  \bibinfo {author} {\bibfnamefont {C.~P.}\ \bibnamefont {Lutz}}, \bibinfo
  {author} {\bibfnamefont {D.~M.}\ \bibnamefont {Eigler}}, \ and\ \bibinfo
  {author} {\bibfnamefont {A.~J.}\ \bibnamefont {Heinrich}},\ }\href {\doibase
  10.1126/science.1214131} {\bibfield  {journal} {\bibinfo  {journal}
  {Science}\ }\textbf {\bibinfo {volume} {335}},\ \bibinfo {pages} {196}
  (\bibinfo {year} {2012})}\BibitemShut {NoStop}%
\bibitem [{\citenamefont {Yan}\ \emph {et~al.}(2017)\citenamefont {Yan},
  \citenamefont {Malavolti}, \citenamefont {Burgess}, \citenamefont
  {Droghetti}, \citenamefont {Rubio},\ and\ \citenamefont {Loth}}]{YAN17}%
  \BibitemOpen
  \bibfield  {author} {\bibinfo {author} {\bibfnamefont {S.}~\bibnamefont
  {Yan}}, \bibinfo {author} {\bibfnamefont {L.}~\bibnamefont {Malavolti}},
  \bibinfo {author} {\bibfnamefont {J.~A.~J.}\ \bibnamefont {Burgess}},
  \bibinfo {author} {\bibfnamefont {A.}~\bibnamefont {Droghetti}}, \bibinfo
  {author} {\bibfnamefont {A.}~\bibnamefont {Rubio}}, \ and\ \bibinfo {author}
  {\bibfnamefont {S.}~\bibnamefont {Loth}},\ }\href {\doibase
  10.1126/sciadv.1603137} {\bibfield  {journal} {\bibinfo  {journal} {Sci.
  Adv.}\ }\textbf {\bibinfo {volume} {3}},\ \bibinfo {pages} {e1603137}
  (\bibinfo {year} {2017})}\BibitemShut {NoStop}%
\bibitem [{\citenamefont {Hirjibehedin}\ \emph {et~al.}(2007)\citenamefont
  {Hirjibehedin}, \citenamefont {Lin}, \citenamefont {Otte}, \citenamefont
  {Ternes}, \citenamefont {Lutz}, \citenamefont {Jones},\ and\ \citenamefont
  {Heinrich}}]{HIRJ07}%
  \BibitemOpen
  \bibfield  {author} {\bibinfo {author} {\bibfnamefont {C.~F.}\ \bibnamefont
  {Hirjibehedin}}, \bibinfo {author} {\bibfnamefont {C.-Y.}\ \bibnamefont
  {Lin}}, \bibinfo {author} {\bibfnamefont {A.~F.}\ \bibnamefont {Otte}},
  \bibinfo {author} {\bibfnamefont {M.}~\bibnamefont {Ternes}}, \bibinfo
  {author} {\bibfnamefont {C.~P.}\ \bibnamefont {Lutz}}, \bibinfo {author}
  {\bibfnamefont {B.~A.}\ \bibnamefont {Jones}}, \ and\ \bibinfo {author}
  {\bibfnamefont {A.~J.}\ \bibnamefont {Heinrich}},\ }\href {\doibase
  10.1126/science.1146110} {\bibfield  {journal} {\bibinfo  {journal}
  {Science}\ }\textbf {\bibinfo {volume} {317}},\ \bibinfo {pages} {1199}
  (\bibinfo {year} {2007})}\BibitemShut {NoStop}%
\bibitem [{\citenamefont {Katsnelson}\ \emph {et~al.}(2000)\citenamefont
  {Katsnelson}, \citenamefont {Dobrovitski},\ and\ \citenamefont
  {Harmon}}]{KATS00}%
  \BibitemOpen
  \bibfield  {author} {\bibinfo {author} {\bibfnamefont {M.~I.}\ \bibnamefont
  {Katsnelson}}, \bibinfo {author} {\bibfnamefont {V.~V.}\ \bibnamefont
  {Dobrovitski}}, \ and\ \bibinfo {author} {\bibfnamefont {B.~N.}\ \bibnamefont
  {Harmon}},\ }\href {\doibase 10.1103/PhysRevA.62.022118} {\bibfield
  {journal} {\bibinfo  {journal} {Phys. Rev. A}\ }\textbf {\bibinfo {volume}
  {62}},\ \bibinfo {pages} {022118} (\bibinfo {year} {2000})}\BibitemShut
  {NoStop}%
\bibitem [{\citenamefont {Hamieh}\ and\ \citenamefont
  {Katsnelson}(2005)}]{HAMI05}%
  \BibitemOpen
  \bibfield  {author} {\bibinfo {author} {\bibfnamefont {S.~D.}\ \bibnamefont
  {Hamieh}}\ and\ \bibinfo {author} {\bibfnamefont {M.~I.}\ \bibnamefont
  {Katsnelson}},\ }\href {\doibase 10.1103/PhysRevA.72.032316} {\bibfield
  {journal} {\bibinfo  {journal} {Phys. Rev. A}\ }\textbf {\bibinfo {volume}
  {72}},\ \bibinfo {pages} {032316} (\bibinfo {year} {2005})}\BibitemShut
  {NoStop}%
\bibitem [{\citenamefont {Donker}\ \emph {et~al.}(2016)\citenamefont {Donker},
  \citenamefont {{De Raedt}},\ and\ \citenamefont {Katsnelson}}]{DONK16}%
  \BibitemOpen
  \bibfield  {author} {\bibinfo {author} {\bibfnamefont {H.~C.}\ \bibnamefont
  {Donker}}, \bibinfo {author} {\bibfnamefont {H.}~\bibnamefont {{De Raedt}}},
  \ and\ \bibinfo {author} {\bibfnamefont {M.~I.}\ \bibnamefont {Katsnelson}},\
  }\href {\doibase 10.1103/PhysRevB.93.184426} {\bibfield  {journal} {\bibinfo
  {journal} {Phys. Rev. B}\ }\textbf {\bibinfo {volume} {93}},\ \bibinfo
  {pages} {184426} (\bibinfo {year} {2016})}\BibitemShut {NoStop}%
\bibitem [{\citenamefont {Katsnelson}\ \emph {et~al.}(2003)\citenamefont
  {Katsnelson}, \citenamefont {Dobrovitski}, \citenamefont {{De Raedt}},\ and\
  \citenamefont {Harmon}}]{KATS03}%
  \BibitemOpen
  \bibfield  {author} {\bibinfo {author} {\bibfnamefont {M.~I.}\ \bibnamefont
  {Katsnelson}}, \bibinfo {author} {\bibfnamefont {V.~V.}\ \bibnamefont
  {Dobrovitski}}, \bibinfo {author} {\bibfnamefont {H.~A.}\ \bibnamefont {{De
  Raedt}}}, \ and\ \bibinfo {author} {\bibfnamefont {B.~N.}\ \bibnamefont
  {Harmon}},\ }\href {\doibase 10.1016/j.physleta.2003.08.046} {\bibfield
  {journal} {\bibinfo  {journal} {Phys. Lett. A}\ }\textbf {\bibinfo {volume}
  {318}},\ \bibinfo {pages} {445} (\bibinfo {year} {2003})}\BibitemShut
  {NoStop}%
\bibitem [{\citenamefont {Katsnelson}\ and\ \citenamefont
  {Kogan}(2005)}]{KATS05}%
  \BibitemOpen
  \bibfield  {author} {\bibinfo {author} {\bibfnamefont {M.~I.}\ \bibnamefont
  {Katsnelson}}\ and\ \bibinfo {author} {\bibfnamefont {E.}~\bibnamefont
  {Kogan}},\ }\href {\doibase 10.1103/PhysRevB.72.104406} {\bibfield  {journal}
  {\bibinfo  {journal} {Phys. Rev. B}\ }\textbf {\bibinfo {volume} {72}},\
  \bibinfo {pages} {104406} (\bibinfo {year} {2005})}\BibitemShut {NoStop}%
\bibitem [{\citenamefont {Katsnelson}\ \emph {et~al.}(2001)\citenamefont
  {Katsnelson}, \citenamefont {Dobrovitski},\ and\ \citenamefont
  {Harmon}}]{KATS01}%
  \BibitemOpen
  \bibfield  {author} {\bibinfo {author} {\bibfnamefont {M.~I.}\ \bibnamefont
  {Katsnelson}}, \bibinfo {author} {\bibfnamefont {V.~V.}\ \bibnamefont
  {Dobrovitski}}, \ and\ \bibinfo {author} {\bibfnamefont {B.~N.}\ \bibnamefont
  {Harmon}},\ }\href {\doibase 10.1103/PhysRevB.63.212404} {\bibfield
  {journal} {\bibinfo  {journal} {Phys. Rev. B}\ }\textbf {\bibinfo {volume}
  {63}},\ \bibinfo {pages} {212404} (\bibinfo {year} {2001})}\BibitemShut
  {NoStop}%
\bibitem [{\citenamefont {Yuan}\ \emph {et~al.}(2007)\citenamefont {Yuan},
  \citenamefont {Katsnelson},\ and\ \citenamefont {{De Raedt}}}]{YUAN07}%
  \BibitemOpen
  \bibfield  {author} {\bibinfo {author} {\bibfnamefont {S.}~\bibnamefont
  {Yuan}}, \bibinfo {author} {\bibfnamefont {M.~I.}\ \bibnamefont
  {Katsnelson}}, \ and\ \bibinfo {author} {\bibfnamefont {H.}~\bibnamefont {{De
  Raedt}}},\ }\href {\doibase 10.1103/PhysRevA.75.052109} {\bibfield  {journal}
  {\bibinfo  {journal} {Phys. Rev. A}\ }\textbf {\bibinfo {volume} {75}},\
  \bibinfo {pages} {052109} (\bibinfo {year} {2007})}\BibitemShut {NoStop}%
\bibitem [{\citenamefont {Yuan}\ \emph {et~al.}(2006)\citenamefont {Yuan},
  \citenamefont {Katsnelson},\ and\ \citenamefont {{De Raedt}}}]{YUAN06}%
  \BibitemOpen
  \bibfield  {author} {\bibinfo {author} {\bibfnamefont {S.}~\bibnamefont
  {Yuan}}, \bibinfo {author} {\bibfnamefont {M.~I.}\ \bibnamefont
  {Katsnelson}}, \ and\ \bibinfo {author} {\bibfnamefont {H.}~\bibnamefont {{De
  Raedt}}},\ }\href {\doibase 10.1134/S0021364006140128} {\bibfield  {journal}
  {\bibinfo  {journal} {JETP Lett.}\ }\textbf {\bibinfo {volume} {84}},\
  \bibinfo {pages} {99} (\bibinfo {year} {2006})}\BibitemShut {NoStop}%
\bibitem [{\citenamefont {Box}\ and\ \citenamefont {Muller}(1958)}]{BOX58}%
  \BibitemOpen
  \bibfield  {author} {\bibinfo {author} {\bibfnamefont {G.~E.~P.}\
  \bibnamefont {Box}}\ and\ \bibinfo {author} {\bibfnamefont {M.~E.}\
  \bibnamefont {Muller}},\ }\href {\doibase 10.1214/aoms/1177706645} {\bibfield
   {journal} {\bibinfo  {journal} {Ann. Math. Statist.}\ }\textbf {\bibinfo
  {volume} {29}},\ \bibinfo {pages} {610} (\bibinfo {year} {1958})}\BibitemShut
  {NoStop}%
\bibitem [{\citenamefont {Hams}\ and\ \citenamefont {{De
  Raedt}}(2000)}]{HAMS00}%
  \BibitemOpen
  \bibfield  {author} {\bibinfo {author} {\bibfnamefont {A.}~\bibnamefont
  {Hams}}\ and\ \bibinfo {author} {\bibfnamefont {H.}~\bibnamefont {{De
  Raedt}}},\ }\href {\doibase 10.1103/PhysRevE.62.4365} {\bibfield  {journal}
  {\bibinfo  {journal} {Phys. Rev. E}\ }\textbf {\bibinfo {volume} {62}},\
  \bibinfo {pages} {4365} (\bibinfo {year} {2000})}\BibitemShut {NoStop}%
\bibitem [{\citenamefont {Dobrovitski}\ and\ \citenamefont {{De
  Raedt}}(2003)}]{DOBR03}%
  \BibitemOpen
  \bibfield  {author} {\bibinfo {author} {\bibfnamefont {V.~V.}\ \bibnamefont
  {Dobrovitski}}\ and\ \bibinfo {author} {\bibfnamefont {H.~A.}\ \bibnamefont
  {{De Raedt}}},\ }\href {\doibase 10.1103/PhysRevE.67.056702} {\bibfield
  {journal} {\bibinfo  {journal} {Phys. Rev. E}\ }\textbf {\bibinfo {volume}
  {67}},\ \bibinfo {pages} {056702} (\bibinfo {year} {2003})}\BibitemShut
  {NoStop}%
\bibitem [{\citenamefont {{De Raedt}}\ and\ \citenamefont
  {Michielsen}(2006)}]{RAED06}%
  \BibitemOpen
  \bibfield  {author} {\bibinfo {author} {\bibfnamefont {H.}~\bibnamefont {{De
  Raedt}}}\ and\ \bibinfo {author} {\bibfnamefont {K.}~\bibnamefont
  {Michielsen}},\ }\enquote {\bibinfo {title} {Computational methods for
  simulating quantum computers},}\ in\ \href@noop {} {\emph {\bibinfo
  {booktitle} {Quantum and molecular computing, quantum simulations}}},\
  \bibinfo {series} {Handbook of Theoretical and Computational Nanotechnology},
  Vol.~\bibinfo {volume} {3},\ \bibinfo {editor} {edited by\ \bibinfo {editor}
  {\bibfnamefont {M.}~\bibnamefont {Rieth}}\ and\ \bibinfo {editor}
  {\bibfnamefont {W.}~\bibnamefont {Schommers}}}\ (\bibinfo  {publisher}
  {American Scientific Publishers},\ \bibinfo {year} {2006})\ Chap.~\bibinfo
  {chapter} {1}, pp.\ \bibinfo {pages} {2--48}\BibitemShut {NoStop}%
\bibitem [{\citenamefont {Fano}(1957)}]{FANO57}%
  \BibitemOpen
  \bibfield  {author} {\bibinfo {author} {\bibfnamefont {U.}~\bibnamefont
  {Fano}},\ }\href {\doibase 10.1103/RevModPhys.29.74} {\bibfield  {journal}
  {\bibinfo  {journal} {Rev. Mod. Phys.}\ }\textbf {\bibinfo {volume} {29}},\
  \bibinfo {pages} {74} (\bibinfo {year} {1957})}\BibitemShut {NoStop}%
\bibitem [{\citenamefont {Misra}(1977)}]{MISR77}%
  \BibitemOpen
  \bibfield  {author} {\bibinfo {author} {\bibfnamefont {E.~C.~G.}\
  \bibnamefont {Misra}, \bibfnamefont {B.and~Sudarshan}},\ }\href {\doibase
  10.1063/1.523304} {\bibfield  {journal} {\bibinfo  {journal} {J. Math.
  Phys.}\ }\textbf {\bibinfo {volume} {18}},\ \bibinfo {pages} {756} (\bibinfo
  {year} {1977})}\BibitemShut {NoStop}%
\bibitem [{\citenamefont {Suzuki}(1976)}]{SUZU76}%
  \BibitemOpen
  \bibfield  {author} {\bibinfo {author} {\bibfnamefont {M.}~\bibnamefont
  {Suzuki}},\ }\href {\doibase 10.1007/BF01609348} {\bibfield  {journal}
  {\bibinfo  {journal} {Commun. Math. Phys.}\ }\textbf {\bibinfo {volume}
  {51}},\ \bibinfo {pages} {183} (\bibinfo {year} {1976})}\BibitemShut
  {NoStop}%
\bibitem [{\citenamefont {Sternberg}(2004)}]{STERN04}%
  \BibitemOpen
  \bibfield  {author} {\bibinfo {author} {\bibfnamefont {S.}~\bibnamefont
  {Sternberg}},\ }\href@noop {} {\emph {\bibinfo {title} {Lie algebras}}}\
  (\bibinfo  {publisher} {http://www.math.harvard.edu/~shlomo/},\ \bibinfo
  {year} {2004})\BibitemShut {NoStop}%
\bibitem [{\citenamefont {{Von Neumann}}(1955)}]{NEUM55}%
  \BibitemOpen
  \bibfield  {author} {\bibinfo {author} {\bibfnamefont {J.}~\bibnamefont {{Von
  Neumann}}},\ }\href@noop {} {\emph {\bibinfo {title} {Mathematical
  Foundations of Quantum Mechanics}}}\ (\bibinfo  {publisher} {Princeton
  University Press},\ \bibinfo {address} {Princeton},\ \bibinfo {year}
  {1955})\BibitemShut {NoStop}%
\bibitem [{\citenamefont {Capriotti}(2001)}]{CAPR01}%
  \BibitemOpen
  \bibfield  {author} {\bibinfo {author} {\bibfnamefont {L.}~\bibnamefont
  {Capriotti}},\ }\href {\doibase 10.1142/S0217979201004605} {\bibfield
  {journal} {\bibinfo  {journal} {Int. J. Mod. Phys. B}\ }\textbf {\bibinfo
  {volume} {15}},\ \bibinfo {pages} {1799} (\bibinfo {year}
  {2001})}\BibitemShut {NoStop}%
\bibitem [{\citenamefont {{van Wezel}}\ \emph {et~al.}(2006)\citenamefont {{van
  Wezel}}, \citenamefont {Zaanen},\ and\ \citenamefont {{van den
  Brink}}}]{WEZE06}%
  \BibitemOpen
  \bibfield  {author} {\bibinfo {author} {\bibfnamefont {J.}~\bibnamefont {{van
  Wezel}}}, \bibinfo {author} {\bibfnamefont {J.}~\bibnamefont {Zaanen}}, \
  and\ \bibinfo {author} {\bibfnamefont {J.}~\bibnamefont {{van den Brink}}},\
  }\href {\doibase 10.1103/PhysRevB.74.094430} {\bibfield  {journal} {\bibinfo
  {journal} {Phys. Rev. B}\ }\textbf {\bibinfo {volume} {74}},\ \bibinfo
  {pages} {094430} (\bibinfo {year} {2006})}\BibitemShut {NoStop}%
\bibitem [{\citenamefont {Fouet}\ \emph {et~al.}(2001)\citenamefont {Fouet},
  \citenamefont {Sindzingre},\ and\ \citenamefont {Lhuillier}}]{FOUE01}%
  \BibitemOpen
  \bibfield  {author} {\bibinfo {author} {\bibfnamefont {J.~B.}\ \bibnamefont
  {Fouet}}, \bibinfo {author} {\bibfnamefont {P.}~\bibnamefont {Sindzingre}}, \
  and\ \bibinfo {author} {\bibfnamefont {C.}~\bibnamefont {Lhuillier}},\ }\href
  {\doibase 10.1007/s100510170273} {\bibfield  {journal} {\bibinfo  {journal}
  {Eur. Phys. J. B}\ }\textbf {\bibinfo {volume} {20}},\ \bibinfo {pages} {241}
  (\bibinfo {year} {2001})}\BibitemShut {NoStop}%
\bibitem [{\citenamefont {Zhu}(2005)}]{ZHU05}%
  \BibitemOpen
  \bibfield  {author} {\bibinfo {author} {\bibfnamefont {Y.}~\bibnamefont
  {Zhu}},\ }\href@noop {} {\emph {\bibinfo {title} {Modern Techniques for
  Characterizing Magnetic Materials}}},\ Springer ebook collection / Chemistry
  and Materials Science 2005-2008\ (\bibinfo  {publisher} {Springer US},\
  \bibinfo {year} {2005})\BibitemShut {NoStop}%
\bibitem [{\citenamefont {Paz}\ and\ \citenamefont {Zurek}(1999)}]{PAZ99}%
  \BibitemOpen
  \bibfield  {author} {\bibinfo {author} {\bibfnamefont {J.~P.}\ \bibnamefont
  {Paz}}\ and\ \bibinfo {author} {\bibfnamefont {W.~H.}\ \bibnamefont
  {Zurek}},\ }\href {\doibase 10.1103/PhysRevLett.82.5181} {\bibfield
  {journal} {\bibinfo  {journal} {Phys. Rev. Lett.}\ }\textbf {\bibinfo
  {volume} {82}},\ \bibinfo {pages} {5181} (\bibinfo {year}
  {1999})}\BibitemShut {NoStop}%
\bibitem [{\citenamefont {Donker}\ \emph {et~al.}(2017)\citenamefont {Donker},
  \citenamefont {{De Raedt}},\ and\ \citenamefont {Katsnelson}}]{DONK17}%
  \BibitemOpen
  \bibfield  {author} {\bibinfo {author} {\bibfnamefont {H.~C.}\ \bibnamefont
  {Donker}}, \bibinfo {author} {\bibfnamefont {H.}~\bibnamefont {{De Raedt}}},
  \ and\ \bibinfo {author} {\bibfnamefont {M.~I.}\ \bibnamefont {Katsnelson}},\
  }\href {\doibase 10.21468/SciPostPhys} {\bibfield  {journal} {\bibinfo
  {journal} {SciPost Phys.}\ }\textbf {\bibinfo {volume} {2}},\ \bibinfo
  {pages} {010} (\bibinfo {year} {2017})}\BibitemShut {NoStop}%
\bibitem [{\citenamefont {Hirjibehedin}\ \emph {et~al.}(2006)\citenamefont
  {Hirjibehedin}, \citenamefont {Lutz},\ and\ \citenamefont
  {Heinrich}}]{HIRJ06}%
  \BibitemOpen
  \bibfield  {author} {\bibinfo {author} {\bibfnamefont {C.~F.}\ \bibnamefont
  {Hirjibehedin}}, \bibinfo {author} {\bibfnamefont {C.~P.}\ \bibnamefont
  {Lutz}}, \ and\ \bibinfo {author} {\bibfnamefont {A.~J.}\ \bibnamefont
  {Heinrich}},\ }\href {\doibase 10.1126/science.1125398} {\bibfield  {journal}
  {\bibinfo  {journal} {Science}\ }\textbf {\bibinfo {volume} {312}},\ \bibinfo
  {pages} {1021} (\bibinfo {year} {2006})}\BibitemShut {NoStop}%
\bibitem [{\citenamefont {Bogoliubov}\ \emph {et~al.}(1986)\citenamefont
  {Bogoliubov}, \citenamefont {Izergin},\ and\ \citenamefont
  {Korepin}}]{BOGO86}%
  \BibitemOpen
  \bibfield  {author} {\bibinfo {author} {\bibfnamefont {N.~M.}\ \bibnamefont
  {Bogoliubov}}, \bibinfo {author} {\bibfnamefont {A.~G.}\ \bibnamefont
  {Izergin}}, \ and\ \bibinfo {author} {\bibfnamefont {V.~E.}\ \bibnamefont
  {Korepin}},\ }\href {\doibase 10.1016/0550-3213(86)90579-1} {\bibfield
  {journal} {\bibinfo  {journal} {Nucl. Phys. B}\ }\textbf {\bibinfo {volume}
  {275}},\ \bibinfo {pages} {687} (\bibinfo {year} {1986})}\BibitemShut
  {NoStop}%
\bibitem [{\citenamefont {Parkinson}\ and\ \citenamefont
  {Farnell}(2010)}]{PARK10}%
  \BibitemOpen
  \bibfield  {author} {\bibinfo {author} {\bibfnamefont {J.~B.}\ \bibnamefont
  {Parkinson}}\ and\ \bibinfo {author} {\bibfnamefont {D.~J.~J.}\ \bibnamefont
  {Farnell}},\ }\href@noop {} {\emph {\bibinfo {title} {{An Introduction to
  Quantum Spin Systems}}}},\ Vol.\ \bibinfo {volume} {816}\ (\bibinfo
  {publisher} {Springer},\ \bibinfo {address} {Berlin Heidelberg},\ \bibinfo
  {year} {2010})\BibitemShut {NoStop}%
\bibitem [{\citenamefont {Hanson}\ \emph {et~al.}(2008)\citenamefont {Hanson},
  \citenamefont {Dobrovitski}, \citenamefont {Feiguin}, \citenamefont {Gywat},\
  and\ \citenamefont {Awschalom}}]{HANS08}%
  \BibitemOpen
  \bibfield  {author} {\bibinfo {author} {\bibfnamefont {R.}~\bibnamefont
  {Hanson}}, \bibinfo {author} {\bibfnamefont {V.~V.}\ \bibnamefont
  {Dobrovitski}}, \bibinfo {author} {\bibfnamefont {A.~E.}\ \bibnamefont
  {Feiguin}}, \bibinfo {author} {\bibfnamefont {O.}~\bibnamefont {Gywat}}, \
  and\ \bibinfo {author} {\bibfnamefont {D.~D.}\ \bibnamefont {Awschalom}},\
  }\href {\doibase 10.1126/science.1155400} {\bibfield  {journal} {\bibinfo
  {journal} {Science}\ }\textbf {\bibinfo {volume} {320}},\ \bibinfo {pages}
  {352} (\bibinfo {year} {2008})}\BibitemShut {NoStop}%
\end{thebibliography}%
\end{document}